\documentclass[a4paper,11pt]{article}
\usepackage{amsmath,amssymb,amsthm}
\usepackage{dsfont}
\usepackage{graphicx}
\usepackage[british]{babel}
\usepackage[utf8]{inputenc}
\usepackage{xcolor}
\usepackage[colorlinks=true]{hyperref}
\usepackage{authblk}
\usepackage{blindtext}
\usepackage{epstopdf}
\usepackage{cite}

\newtheorem{remark}{Remark}

\newcommand{\R}{\mathbb{R}}
\newcommand{\be}{\begin{equation}}
\newcommand{\ee}{\end{equation}}
\newcommand{\fer}[1]{(\ref{#1})}

\newcommand{\rev}[1]{\textcolor{black}{#1}}

\def\e{\epsilon}

\newenvironment{equations}{\equation\aligned}{\endaligned\endequation}
\begin{document}
\title{\rev{Kinetic models for epidemic dynamics with social heterogeneity }}


\author[1]{G. Dimarco\thanks{\tt giacomo.dimarco@unife.it}}
\author[2]{B. Perthame\thanks{\tt benoit.perthame@ljll.math.upmc.fr}}
\author[3]{G. Toscani\thanks{\tt giuseppe.toscani@unipv.it}}
\author[3]{M. Zanella\thanks{\tt mattia.zanella@unipv.it}}

\affil[1]{\normalsize
        Mathematics and Computer Science Department,
       
        University of Ferrara, Italy.}
 \affil[2]{\normalsize Sorbonne Universit\'e, CNRS, Universit\'e de Paris,  Inria, 
 
 Laboratoire Jacques-Louis Lions, 75005 Paris, France.}
\affil[3]{
        Mathematics Department, 
        University of Pavia, Italy.}
\date{}

\maketitle

\abstract{
\rev{We introduce a mathematical description of the impact of sociality in the spread of infectious diseases by integrating an epidemiological dynamics with a kinetic modeling of population-based contacts.} The kinetic description leads to study the evolution over time of Boltzmann-type equations describing the number densities of social contacts of susceptible, infected and recovered individuals, whose proportions are driven by a classical SIR-type compartmental model in epidemiology. Explicit calculations show that the spread of the disease is closely related to moments of the contact distribution.  Furthermore, the kinetic model allows to clarify how a selective control can be assumed to achieve a minimal lockdown strategy by only reducing individuals undergoing a very large number of daily contacts. 
We conduct numerical simulations which confirm the ability of the model to describe different phenomena characteristic of the rapid spread of an epidemic. Motivated by the COVID-19 pandemic, a last part is dedicated to fit numerical solutions of the proposed model with infection data coming from different European countries.
}


\section{Introduction}

\rev{The SARS-CoV-2 pandemic} led in many countries to heavy lockdown measures assumed by the governments with the aim to control and limit its \rev{spreading.}
\rev{In this context, an essential role is played by the mathematical modeling of infectious diseases since they allow direct validation with real data, unlike other classical phenomenological approaches.} This consequently permits evaluation of control and prevention strategies by comparing their cost with effectiveness and to give support to public health decisions \cite{Ferg, Ril}. On this subject, most of the models present in literature make assumptions on transmission parameters \cite{BCF, DH} which are considered the only responsible of the spread of the infection. However, special attention was recently paid by the scientific community to the role and the estimate of the distribution of contacts between individuals as also a relevant cause of the potential pathogen transmission (cf. \cite{Plos,DolT,Fuma,Moss} and the references therein). 

In this direction, the results reported in \cite{Plos} can be of great help when designing partial lockdown strategies. In fact, an optimal control of the pathogen transmission of the epidemic could be achieved through a direct limitation of the number of daily contacts among people. 
On this subject, the detailed analysis performed in \cite{Plos} put into evidence that the number of social contacts in the population is in general well-fitted by a Gamma distribution, even if this distribution is not uniform with respect to age, sex and wealth. Gamma distributions belong to the wide class of generalized Gamma distributions \cite{Lie, Sta}, which have been recently connected to the statistical study of social phenomena \cite{KK, Reh}, and fruitfully described as steady states of kinetic models  aiming to describe the formation of these profiles in consequence of repeated elementary interactions  \cite{DT, To3}.

Starting from the above consideration and inspired by the recent development concerning kinetic models describing human behavior \cite{DT,To3}, in this paper we develop a mathematical framework to connect the distribution of social contacts with the spreading of a disease in a multiagent system. This result is obtained by integrating an epidemiological dynamics given by a classical compartmental model \cite{BCF, DH, HWH00} with an approach based on kinetic equations determining the formation of social contacts. In this paper we concentrate on the classical SIR dynamics. However, we stress that the ideas here described are clearly not reduced to this model which can be intended as an example. Instead, the methodology here discussed can be extended to incorporate more realistic epidemiological dynamics, like for instance the classical endemic models discussed in \cite{BCF, DH, HWH00}. \rev{In particular,  the extension of the present approach to age-dependent compartmental models could be of great interest to produce realistic scenarios. }

 Other aspects, which certainly have a stronger impact on how a virus spreads, are related to the presence of asymptomatic individuals~\cite{Gaeta} as well as to a time delay between contacts and outbreak of the disease~\cite{Cook} which helps in the diffusion of the illness. These possible modeling improvements are the subject of future investigations and they will not be treated in this work. 
  \rev{In fact, we stress that the principal scope of this work is to introduce a new class of models which is capable to incorporate information on the social heterogeneity of a population which we believe to be a crucial aspect in the spread of contagious diseases.}
 Besides \rev{these} simplifying assumptions, we will show that the basic features considered and detailed in the rest of the article are sufficient in many cases to construct a \rev{new class of models which well fits with the experimental data. Precise quantitative estimates are postponed to future investigations.}  

An easy way to understand epidemiology models is that, given a population composed of agents, they prescribe movements of individuals between different states based on some matching functions or laws of motion. According to the classical SIR models \cite{HWH00}, agents in the system are split into three classes, the susceptible, who can contract the disease, the infected, who have already contracted the disease and can transmit it, and the recovered who are healed, immune or isolated. Inspired by the model considered in \cite{DT} for describing a social attitude and making use of classical epidemiological dynamics, we present here a model composed by a system of three kinetic equations, each one describing the time evolution of the distribution of the number of contacts for the subpopulation belonging to a given epidemiological class. These three equations are further coupled by taking into account the movements of agents from one class to the other as a consequence of the infection, \rev{with an intensity proportional to the product of the average contact frequencies, rather than the product of population fractions}. 

The interactions which describe the social contacts of individuals are based on few rules and can be easily adapted to take into account the behavior of agents in the different classes in presence of the infectious disease. Our joint framework is consequently based on two models which can be considered classical in the respective fields. 

From the side of multi-agent systems in statistical mechanics, the linear kinetic model introduced in \cite{DT, To3} has been shown to be flexible and able to describe, with \rev{suitable} modifications, different problems in which human behavior plays an essential role, like the formation of social contacts. \rev{Once the statistical distribution of social contacts has been properly identified as equilibrium density of the underlying kinetic model, this information is used to close the hierarchy of equations describing the evolution of moments \cite{Cer}. In this way, we obtain a coupled system of equations, identifying a \rev{new epidemiological} model which takes into account at best the statistical details about the contact distribution of a population. 
 The  model connects the measure of the heterogeneity of the population, i.e. the variance of the contact distribution, with the epidemic trajectory. This is in agreement with a well-know finding in the epidemiological literature, see e.g. \cite{AM85,Diek, Novo, Van}. A recent research showing the influence of population heterogeneity on herd immunity to COVID-19 infection is due to Britton et al. \cite{BBT}.}

\rev{Starting from the general macroscopic model, one can fruitfully obtain from it various sub-classes of SIR-type epidemiological models characterized by non-linear incidence rates}, as for instance recently considered in \cite{KM}. It is also interesting to remark that the presence of non-linearity in the incidence rate function, and in particular, the concavity condition with respect to the number of infected has been considered in \cite{KM} as a consequence of \emph{psychological} effects. Namely, the authors observed that in the presence of a very large number of infected, the probability for an infected to transmit the virus may further decrease because the population \emph{tend to naturally reduce the number of contacts}. 
The importance of reducing at best the social contacts to countering the advance of a pandemic is a well-known and  studied phenomenon \cite{Ferg}. While in normal life activity, it is commonly assumed that a large part of \rev{agents behave} in a similar way, in presence of an extraordinary situation like the one due to a pandemic, it is highly reasonable to conjecture that the social behavior of individuals is strictly affected by their personal feeling in terms of safeness. Thus, in this work, we focus on the assumption that it is the degree of diffusion of the disease that changes people's behavior in terms of social contacts, in view both of the personal perception and/or of the external government intervention. More generally, the model can be clearly extended to consider more realistic dependencies between an epidemic disease and the social contacts of individuals. However, this does not change the essential conclusions of our analysis, namely that there is a close interplay between the spread of the disease and the distribution of contacts, that the kinetic description is able to quantify. \rev{ In particular, we stress the fact that we consider our approach as methodological}, thus the encouraging results described in the rest of the article suggest that a similar analysis can be carried out, at the price of an increasing difficulty in computations, in more complex epidemiological models like the SIDARTHE model \cite{Bruno,Gatto}, to validate and improve the eventual partial lockdown strategies of the government and to suggest future measures.

The rest of the paper is organized as follows. Section~\ref{model} introduces the system of three SIR-type kinetic equations combining the dynamics of social contacts with the spread of infectious disease in a system of interacting individuals. Then, in Section~\ref{FP-limit} we show that through a suitable asymptotic procedure, the solution to the kinetic system tends towards the solution of a system of three SIR-type Fokker-Planck type equations with local equilibria of Gamma-type~\cite{Plos}. Once the system of Fokker-Planck type equations has been derived, in Section~\ref{splitting}, we close the system of kinetic equations around the Gamma-type equilibria to obtain \rev{a new epidemiological model in which the incidence rate depends on the number of social contacts between individuals}. Last, in Section~\ref{numerics}, we investigate at a numerical level the relationships between the solutions of the kinetic system of Boltzmann type, its Fokker-Planck asymptotics and the macroscopic model. These simulations confirm the ability of our approach to describe different phenomena characteristic of the trend of social contacts in situations compromised by the rapid spread of an epidemic and the consequences of various lockdown action in its evolution. \rev{A last part is dedicated to a fitting of the model with the experimental observations: first we estimate the parameters of the epidemic through the data at disposal and successively we use them in the macroscopic model showing that our approach is able to reproduce the pandemic trend.}

\section{\rev{A kinetic approach combining social contacts and epidemic dynamics}}\label{model}

Our goal is to build a kinetic system which suitably describes the spreading of an infectious disease under \rev{the} dependence of the contagiousness parameters on the individual number of social contacts of the agents. Accordingly to classical SIR models \cite{HWH00}, the entire population is divided into three classes: susceptible, infected and recovered individuals. \rev{As already claimed the ideas here developed can be extended to more complex compartmental epidemic models.}

Aiming to understand social contacts effects on the dynamics, we will not consider in the sequel the role of \rev{other sources of possible} heterogeneity in the disease parameters (such as the personal susceptibility to a given disease), which could be derived from the classical epidemiological models, suitably adjusted to account for new information \cite{Diek, Novo, Van}. Consequently, agents in the system are considered indistinguishable \cite{PT13}. This means that the state of an individual in each class at any instant of time $t\ge 0$ is completely characterized by the sole number of contacts $x \ge0$, measured in some unit.

While $x$ is a natural positive number at the individual level, without loss of generality we will consider $x$ in the rest of the paper to be a nonnegative real number, $x\in \mathbb{R}_+$, at the population level. We denote then by $f_S(x,t)$, $f_I(x,t)$ and $f_R(x,t)$, the distributions at time $t > 0$ of the number of social contacts of the population of susceptible, infected and recovered individuals, respectively. The distribution of contacts of the whole population is then recovered as the sum of the three distributions
\[
f(x,t)=f_S(x,t)+f_I(x,t)+f_R(x,t).
\]
We do not consider for simplicity of presentation disease related mortality as well as the presence of asymptomatic individuals \rev{which we aim to insert in future investigations}. Therefore, we can fix the total distribution of social contacts to be a probability density for all times $t \ge 0$
\[
\int_{\mathbb{R}_+} f(x,t)\,dx = 1.
\]
As a consequence, the quantities 
\begin{equation}\label{mass}
J(t)=\int_{\mathbb{R}^+}f_J(x,t)\,dx,\quad J \in \{S,I,R\}  
\end{equation}
denote the fractions, at time $t \ge 0$, of susceptible, infected and recovered respectively. For a given constant $\alpha>0$, and time $t \ge 0$, we also denote with $x^\alpha(t)$ the moment of the number of contacts distribution $f(x,t)$ of order $\alpha$ (divided by the unitary mass)
\rev{\[
x_\alpha(t) =\frac{\int_{\mathbb{R}^+}x^\alpha\, f(x,t)\,dx}{\int_{\mathbb{R}^+} f(x,t)\,dx}.
\]}
\rev{In the same way, we denote with $x_{J,\alpha}(t)$ the local} moments of order $\alpha$ for the distributions of the number of contacts in each class \rev{conveniently divided by the mass of the class} 
\be\label{means}
x_{J,\alpha}(t)= \frac 1{\rev{J(t)}}\int_{\mathbb{R}^+}x^\alpha f_J(x,t)\,dx, \quad J \in \{S,I,R\}.
\ee
Unambiguously, we will indicate the mean \rev{and the local mean} values, corresponding to $\alpha=1$, by $x(t)$ and, respectively, $x_J(t)$, $J \in \{ S,I,R\}$.

In what follows, we assume that 
the various classes in the model could act differently in the social process constituting the contact dynamics. The kinetic model then follows combining the epidemic process with the contact dynamics. 
This gives the system 
\begin{equations}\label{sir-gamma}
\frac{\partial f_S(x,t)}{\partial t} &= -K_\e(f_S,f_I)(x,t) +   Q_S(f_S)(x,t)
\\
\frac{\partial f_I(x,t)}{\partial t} &= K_\e(f_S,f_I)(x,t)  - \gamma_\e f_I(x,t) + Q_I(f_I)(x,t)
\\
\frac{\partial f_R(x,t)}{\partial t} &= \gamma_\e f_I(x,t) +  Q_R(f_R)(x,t) 
\end{equations}
where $\gamma_\e$ is the constant recovery rate while the transmission of the infection is governed by the function
 $K_\e(f_S,f_I)$, the local incidence rate, expressed by
 \be\label{inci}
 {K_\e(f_S,f_I)(x, t) = f_S(x,t) \int_{\R^+} \kappa_\e(x,y)f_I(y,t) \,dy.}
 \ee
\rev{In full generality, we will assume that both the recovery rate $\gamma$ and the contact function $\kappa$  depend on a small positive parameter $\e \ll1$ which measures their intensity.}
In \fer{inci} the contact function $\kappa_\e(x,y)$ is a nonnegative {function growing with respect to the number of contacts $x$ and $y$ of the populations of susceptible and infected, and such that $\kappa_\e(x, 0) = 0$}. A leading example for $\kappa_\e(x,y)$ is obtained by choosing
\[
\kappa_\e(x,y) = \beta_\e\, x^\alpha y^\alpha,
\]
where $\alpha, \beta_\e$ are positive constants, that is by taking the incidence rate \rev{dependent on the} product of the number of contacts of susceptible and infected people. When $\alpha=1$, for example, the incidence rate takes the simpler form
\be\label{simple}
{K_\e(f_S,f_I)(x,t) = \beta_\e \, x f_S(x,t) x_I(t)\,I(t).}
\ee
Let observe that with the choices done, the spreading of the epidemic depends heavily on the function $\kappa_\e(\cdot,\cdot)$ used to quantify the rate of possible contagion in terms of the number of social contacts between the classes of susceptible and infected. 

In our combined epidemic contact model \fer{sir-gamma}, the operators $Q_J$, $J\in \{S,I,R\}$ characterize the thermalization of the distribution of social contacts in the various classes. To that aim, let observe that the evolution of the mass fractions $J(t)$, $J\in\{S,I,R\}$ obeys to a classical SIR model by choosing $Q_S \equiv 0$ and $\kappa_\e(x,y) \equiv \beta >0$, thus considering the spreading of the disease independent of the intensity of social contacts. 

\rev{The $Q_J$, $J\in \{S,I,R\}$ 
are integral operators that modify the distribution of contacts $f_J(x,t)$, $J\in \{S,I,R\}$ through repeated interactions among individuals \cite{DT, To3}}. Their action on observable quantities \rev{$\varphi(x)$} is given by  
\be\label{oper}
\int_{\R_+}\varphi(x)\,Q_J(f_J)(x,t)\,dx  = \,\,
  \Big \langle \int_{\R_+}B(x) \bigl( \varphi(x_J^*)-\varphi(x) \bigr) f_J(x,t)
\,dx \Big \rangle.
 \ee
where $B(x)$ measures the interaction frequency, $\langle \cdot \rangle$ denotes mathematical expectation with respect to a random quantity, and $x_J^*$ \rev{denotes the updated} value of the number $x$ of social contacts of the $J$-th population \rev{as a result of an interaction}. \rev{We discuss in the sequel the construction of the social contact model.}

\rev{\subsection{On the distribution of social contacts}}
The process of formation of the distribution of social contacts is obtained by taking into account the typical aspects of human being, in particular the search, in absence of epidemics, of opportunities for socialization. In addition to that, social contacts are due to the common use of public transportations to reach schools, offices and, in general, places of work \rev{as well as} to basic needs of interactions due to work duties. As shown in \cite{Plos}, this leads  individuals to stabilize on a characteristic number of daily contacts depending on the social habits of \rev{a country}.
This quantity is represented in the following by a suitable value $\bar x_M$, which can be considered as the mean number of contacts relative to the population under investigation. \rev{This kind of dynamics and the relative distribution of average daily contacts observed in \cite{Plos} is the one we aim to explain and reproduce in our model.}

\rev{As a final result of our investigation}, we look to a characterization of the distribution of social contacts in a multi-agent system, \rev{the so-called macroscopic behavior}. This can be obtained starting from some universal assumption about the personal behavior of \rev{the single} agents, i.e. from the microscopic behavior. Indeed, as in many other human activities, the daily amount of social contacts is the result of a repeated upgrading based on well-established rules. To this extent, it is enough to recall that the daily life of each person is based on a certain number of activities, and each activity carries a certain number of contacts. Moreover, for any given activity, the number of social contacts varies in consequence of the personal choice. The typical example is the use or not of public transportations to reach the place of work or the social attitudes which scales with the age. Clearly, independently of the personal choices or needs, the number of daily social contacts contains a certain amount of randomness, that has to be taken into account. Also, while it is very easy to reach a high number of social contacts attending crowded places for need or will, going  below a given threshold is very difficult, since various contacts are forced by working or school activities, among others causes. This asymmetry between growth and decrease, as exhaustively discussed in \cite{DT,GT19}, can be suitably modeled by resorting to a so-called \textit{value function} \cite{KT} description of the elementary variation of the $x$ variable measuring the average number of daily contacts. \rev{We will come back to the definition of the value function later in the section.}

\begin{remark} \label{rem:mean}
It is important to outline that, in presence of an epidemic, the characteristic mean number of daily contacts $\bar x_M$ reasonably changes in time, even in absence of an external lockdown intervention, in reason of the perception of danger linked to social contacts. 
Consequently, even if not always explicitly indicated, we will assume  $\bar x_M= \bar x_M(t)$. 
\end{remark}

Furthermore, an important aspect of the formation of the number of social contacts is that their frequency is not uniform with respect to the values of $x$. Indeed, it is reasonable to assume that the frequency of interactions is inversely proportional to the number of contacts $x$. This relationship takes into account that it is highly probable to have at least some contacts, and the rare situation in which one reaches a very high values of contacts \rev{$x\gg\bar x_M$}. \rev{On this subject, we mention a related approach discussed in \cite{FPTT19}. }
The introduction of a variable kernel $B(x)$ into the kinetic equation does not modify the shape of the equilibrium density as shown later, but it allows a better physical description of the phenomenon under study, including an exponential rate of relaxation to equilibrium for the underlying Fokker-Planck type equation \rev{ derived from the kinetic equation that it we will introduced next in \ref{FP-limit}.} 

Following \cite{DT,GT19, To3}, we will now illustrate the mathematical formulation of the previously discussed behavior. In full generality, we assume that individuals in different classes can have a different mean number of contacts. \rev{Then,} the microscopic updates of social contacts of individuals in the class $J\in \{S,I,R\}$ will be taken of the form
 \be\label{coll}
 x_J^* = x  - \Phi^\e (x/\bar x_J) x + \eta_\e x.
 \ee
In a single update (interaction), the number $x$ of contacts can be modified for two reasons, expressed by two terms, both proportional to the value $x$. In the first one, the coefficient $\Phi^\e(\cdot)$, which \rev{takes} both positive and negative values, characterizes the typical and predictable variation of the social contacts of agents, namely the personal social behavior of agents. The second term takes into account a certain amount of unpredictability in the process. \rev{A frequent choice in this setting} consists in assuming that the random variable \rev{$\eta_\e$ is of zero mean and bounded variance of order $\e>0$, expressed by $\langle \eta_\e \rangle =0$, $\langle \eta_\e^2 \rangle  = \e\lambda$, with $\lambda >0$. Furthermore, we assume that $\eta_\e$ has finite moments up to order three. }  
 
The function $\Phi^\e$ plays the role of the \emph{value function} in the prospect theory of Kahneman and \rev{Tversky} \cite{KT, KT1}, and contains the mathematical details of the expected human behavior in the phenomenon under consideration. In particular, the \rev{main} hypothesis on which \rev{this function} is built is that, in relationship with the mean value $\bar x_J$, $J\in \{S,I,R\}$, it is \rev{considered} normally easier to increase the value of $x$ \rev{(individuals look for larger networks)} than to decrease it \rev{(people maintain as much connections as possible)}.
In terms of the variable $ s = x/\bar x_J$ we consider then as in \cite{DT} the class of value functions \rev{obeying to the above general rule} given by
 \be\label{vd}
 \Phi_\delta^\e(s) = \mu \frac{e^{\e(s^\delta -1)/\delta}-1}{e^{\e(s^\delta -1)/\delta}+1 } , \quad  s \ge 0,
 \ee
where the value $\mu$ denotes the maximal amount of variation of $x$ that agents will be able to obtain in a single interaction, $0 < \delta \le 1$ is a suitable constant characterizing the intensity of the individual behavior, while $\e >0$ is related to the intensity of the interaction. Hence, the choice $\e\ll 1$ corresponds to small variations of the mean difference $\langle x_J^* -x\rangle$. Thus, if both effects, randomness and adaptation are scaled with this interaction intensity $\e$, it is possible to equilibrate their effects as we will show in Section~\ref{FP-limit} \rev{and obtain a stationary distribution of contacts}. 
Note also that the value function $\Phi_\delta^\e(s)$ is such that 
\[
  -\mu \le \Phi_\delta^\e(s) \le \mu
  \]
and clearly, the choice $\mu <1$ implies that, in absence of randomness, the value of $x_J^*$ remains positive if $x$ is positive. 

Once the  \emph{elementary interaction}  \fer{coll} is given, for any choice of the value function, the study of the time-evolution of the distribution of the number $x$ of social contacts follows by
resorting to kinetic collision-like approaches \cite{Cer,PT13}, that quantify at any given time the variation of the density of the contact variable $x$ in terms of the interaction operators. 

Thus, for a given density $f_J(x,t)$, $J\in \{S,I,R\}$, we can measure the action on the density of the interaction operators $Q_J(f)(x,t)$  in equations \fer{sir-gamma} fruitfully written in weak form. This form corresponds to say that for all smooth functions $\varphi(x)$ (the observable quantities) we have
 \begin{equation}
  \label{kin-w}
\dfrac{d}{dt} \int_{\R_+}\varphi(x)f_J(x,t)\,dx  = 
  \Big \langle \int_{\R_+}B(x) \bigl( \varphi(x_J^*)-\varphi(x) \bigr) f_J(x,t)
\,dx \Big \rangle.
 \end{equation}
 Here, \rev{the} expectation $\langle \cdot \rangle$ takes into account the presence of the random parameter $\eta_\e$ in the microscopic interaction \fer{coll} while the function $B(x)$, \rev{as already discussed}, measures the interaction frequency.
The right-hand side of equation \fer{kin-w} quantifies the variation in density, at a given time $t>0$, of individuals in the class $J\in \{S,I,R\}$ that modify their value from $x$ to $x_J^* $ (loss term with negative sign) and agents  that  change their value from  $x_J^*$ to $x$  (gain term with positive sign).
\rev{In many situations,} the interaction kernel $B(x)$ can been assumed constant \cite{DT}. This simplifying hypothesis is not always well justified from a modeling point of view and thus in this work, we consider instead a non constant collision kernel $B(x)$ (see \cite{FPTT19} for a discussion on this aspect).
Thus, following the approach in \cite{FPTT19, To3}, we  express the mathematical form of the kernel $B(x)$ by assuming  that the frequency of changes \rev{ in the number of social contacts depends on $x$ itself through the following law}
\[
 B(x) = x^{-b},
 \]
for some constant $b >0$. This kernel assigns a low probability to interactions in which individuals have already a large number of contacts and assigns a high probability to interactions \rev{when}  the value of the variable $x$ is small. 
The constants $a$ and $b$ can be suitably chosen by resorting to the following argument \cite{To3}. For small values of the $x$ variable, the rate of variation of the value function \fer{vd} is given by
 \be\label{gro}
\frac d{dx} \Phi_\delta^\e\left(\frac x{\bar x_J}\right) \approx {\mu^2 \e}\, {\bar x_J^{-\delta}}\, x^{\delta -1}. 
 \ee
Hence, for small values of $x$, the mean individual rate predicted by the value function is proportional to $x^{\delta -1}$. Then, the choice $b = \delta$ would correspond to a collective rate of variation of the system independent of the parameter $\delta$ which instead characterizes the individual rate of variation of the value function.

\rev{In the next section, we investigate the steady states of the interaction operators $Q_J(f)(x,t), \ J\in \{S,I,R\}$ which permit to derive the macroscopic epidemic model containing the effects of social interactions among individuals described in Section \ref{splitting}.}

\subsection{Asymptotic scaling and steady states}\label{FP-limit}
\rev{Let us focus on the sole social contact dynamic and introduce a time scaling 
\[
\tau = \epsilon t, \qquad f_{J,\e}(x,\tau) = f_J(x,\tau/\epsilon), \qquad J \in \{S,I,R\}.
\]
which, in the following, will permit to separate the scale of the epidemic from the time scale at which, by hypothesis, the social contacts acts. 
Then, as a result of the scaling, the distribution $f_{J,\e}$ is solution to 
\begin{equation}
	\label{eq:ki}
	\dfrac{d}{d\tau} \int_{\mathbb R_+} \varphi(x) f_{J,\e}(x,\tau)dx = \dfrac{1}{\e}   \Big \langle \int_{\R_+}B(x) \bigl( \varphi(x_J^*)-\varphi(x) \bigr) f_{J,\e}(x,\tau)
	\,dx \Big \rangle.
\end{equation}}
We concentrate now on the analysis of the asymptotic states of the social contact dynamic in the case in which elementary interactions \fer{coll} produces extremely small modification of the number of social contacts.
To that aim, note that, \rev{from the definition of $\Phi_\delta^\epsilon$ in \eqref{vd} and the assumptions on the noise term $\eta_\e$ we have}
\be\label{ottimo}
\lim_{\e \to 0} \frac 1\e { \Phi_\delta^\e\left(\frac x{\bar x_J}\right)} = \frac\mu{2\delta} \left[\left(\frac x{\bar x_J} \right)^\delta -1\right], \quad \lim_{\e \to 0} \frac 1\e \langle \eta_\e^2\rangle = \lambda.
\ee
Consequently, the actions of both the value function and  the random part of the elementary interaction in \fer{coll} survive in the limit $\e \to 0$. Let observe that the limit procedure induced by \fer{ottimo} corresponds precisely to the situation of small interactions while, at the same time, \rev{the time scale of the dynamics is suitably scaled to see their effects.} In kinetic theory, this is a well-known procedure with the name of \emph{grazing limit}, we point the interested reader to \cite{CPT05, FPTT, PT13} for further details. \rev{Since if $\e\ll1 $ the difference $x^*_J - x$ is small and, assuming $\varphi\in \mathcal C_0$, we can perform the following Taylor expansion
\[
\varphi(x^*_J)-\varphi(x) = (x^*_J - x) \varphi'(x) +  \dfrac{1}{2} (x^*_J-x)^2  \varphi''(x) + \dfrac{1}{6}(x^*_J - x)^3\varphi'''(\hat x_J), 
\]
being $\hat x_J \in (\min\{x,x^*_J\},\max\{x,x^*_J\})$. Writing $x^*_J-x = - \Phi^\e_\delta(x/{\bar x_J})x+x\eta_\e$ from \eqref{coll} and plugging the above expansion in the kinetic model \eqref{eq:ki} we have for $J\in\{S,I,R\}$
\[
\begin{split}
&\dfrac{d}{d\tau} \int_{\mathbb R_+}\varphi(x) f_{J,\e}(x,\tau)dx =  \\
&\qquad\dfrac{1}{\e} \left[ \int_{\mathbb R_+} -\Phi_\delta^\e(x/{\bar{x}_J})x^{1-\delta} \varphi'(x)f_{J,\e}(x,\tau)dx + \dfrac{\lambda\e}{2} \int_{\mathbb R_+} \varphi''(x)x^{2-\delta}f_{J,\e}(x,\tau)dx \right] + R_\varphi(f_{J,\e}),
\end{split}\]
where $R_\varphi(f_{J,\e})$ is the remainder
\[
\begin{split}
R_\varphi(f_{J,\e})(x,\tau) =& \dfrac{1}{2\e}\int_{\mathbb R_+}\varphi''(x) x^{-\delta}(\Phi_\delta^\e(x/{\bar{x}_J})x)^2f_{J,\e}(x,t)dx \\
& \dfrac{1}{6\e} \left\langle  \int_{\mathbb R_+} \varphi'''(\hat x_J) x^{-\delta}(-\Phi^\e_\delta(x/{\bar{x}_J})x + x\eta_\e)^3 f_{J,\e}(x,t)dx \right\rangle . 
\end{split}\]
Since, by assumption, $\varphi$ and its derivatives are bounded in $\mathbb R_+$ and decreasing at infinity and since $\eta_\e$ has bounded moment of order three, namely $\langle |\eta_\e|^3\rangle<+\infty$, using the bound \eqref{ottimo} we can easily argue that in the limit $\e \rightarrow 0^+$ we have
\[
|R_\varphi(f_J)| \rightarrow 0.
\]
Hence, it can be shown that $f_{J,\e}$ converges, up to subsequences, to a distribution function $f_J$ solution to 
\[
\dfrac{d}{d\tau} \int_{\mathbb R_+}\varphi(x)f_{J}(x,\tau)dx =   \int_{\mathbb{R}_+} \left\{-\varphi'(x) \, \frac{\mu \,x^{1-\delta}}{2\delta}\left[\left(\frac x{\bar x_J} \right)^\delta -1\right] +  \frac\lambda{2}\varphi''(x)\,x^{2-\delta} \right\}f_{J}(x,\tau)\,dx
\]
If we also impose at $x=0$ the following no-flux boundary conditions 
\begin{equation}\label{bc}
\frac{\partial}{\partial x} (x^{2-\delta} f_J(x,\tau))\Big|_{x=0} = 0 \quad J\in\{S,I,R\},
\end{equation}
the limit equation coincides with the Fokker-Planck type equation
\[
\dfrac{\partial}{\partial \tau}f_J(x,\tau) = Q^\delta_J(f_J)(x,\tau), 
\]
where
 \be\label{FP-ope}
 Q_J^\delta(f_J)(x,\tau) =  \frac{\mu}{2\delta}\frac{\partial}{\partial x}\left\{\,x^{1-\delta}\left[\left(\frac x{\bar x_J} \right)^\delta -1\right]f_{J}(x,\tau)\right\} +\frac{\lambda}{2} \frac{\partial^2}{\partial x^2} (x^{2-\delta} f_J(x,\tau)), \quad J \in \{S,I,R\}
  \ee
is characterized by a variable diffusion coefficient. }
\rev{Remarkably enough, we can compute explicitly the equilibrium distribution of the surrogate Fokker-Planck model. Indeed, assuming that the mass of the initial distribution is one and the mean values $\bar x_J$, $J \in \{S,I,R\}$ are constant, and by setting $\nu = \mu/\lambda$,  the equilibria are given by the functions
\be\label{equilibrio}
f_J^\infty(x) =  C_J(\bar x_J,\delta,\nu) x^{\nu/\delta +\delta -2}  \exp\left\{ - \frac \nu{\delta^2} \left( \frac x{\bar x_J} \right)^\delta\right\},\qquad  J \in \{S,I,R\},
\ee 
 where $C_J> 0$ is a normalization constant. We may rewrite the obtained steady state \fer{equilibrio}  as a generalized Gamma probability density $f_\infty(x;\theta, \chi,\delta)$ defined by \cite{Lie,Sta}
 \be\label{equili}
f_{J,\infty}(x;\theta, \chi,\delta) = \frac \delta{\theta^\chi} \frac 1{\Gamma\left(\chi/\delta \right)} x^{\chi -1}
\exp\left\{ - \left( x/\theta\right)^\delta\right\},
 \ee 
characterized in terms of the shape $\chi>0$, the scale parameter $\theta >0$, and the exponent $\delta >0$ that in the present situation are given by 
 \be\label{para}
 \chi =  \frac\nu\delta +\delta -1,  \quad  \theta = \bar x_J \left( \frac{\delta^2}\nu\right)^{1/\delta}.
 \ee
 }
It has to be remarked that the shape $\chi$ is positive, only if the constant $\nu = \mu/\lambda$ satisfies the bound
 \be\label{cc}
\nu >\delta(1-\delta).
 \ee
Note that condition \fer{cc} holds, independently of $\delta$, when \rev{$\mu \ge \frac{4}{\lambda}$}, namely when the variance of the random variation in \fer{coll} is small with respect to the maximal variation of the value function.  
Note moreover that for all values $\delta >0$ the moments are expressed in terms of the parameters denoting respectively the mean $\bar x_J$, $J \in \{S,I,R\}$, the variance $\lambda$ of the random effects and the values $ \delta$ and $\mu$ characterizing the value function $\phi_\delta^\e$ defined in \fer{vd}. Finally,
the standard Gamma and Weibull distributions are included in \fer{equilibrio}, and are obtained by choosing $\delta =1$ and, respectively $\delta = \chi$. In both cases, the shape $\chi =\nu$, and no conditions are required for its positivity.
\rev{It is important to notice that the function \eqref{equili} expressing the equilibrium distribution of the daily social contacts in a society is in agreement with the one observed in \cite{Plos}. This was one of the goals of our investigation.} 

\section{\rev{The macroscopic social-SIR dynamics}}\label{splitting}
Once we have obtained the characterization of the equilibrium distribution of the transition operators $Q_J$, $J \in \{S,I,R\}$, we are ready to study the complete system \eqref{sir-gamma}. In the rest of this section we will determine the observable macroscopic equations of the introduced kinetic model. 
 
\subsection{\rev{Derivation of moment based systems}}\label{ssir}
\rev{
The assumption that the dynamics leading to the contact formation is much faster than the epidemic dynamics corresponds to consider $ \beta_\e = \epsilon \beta$ in \eqref{inci} and $\gamma_\e  = \e\gamma$ with $\beta,\gamma>0$, being $\e\ll1$ the scaling parameter introduced in the previous section. After introducing the scaled distributions $f_J(x,\tau) = f_J(x,\tau/\e)$ and noticing that $\frac{\partial}{\partial \tau} f_J = \frac{1}{\e}\frac{\partial}{\partial t }f_J$} we can rewrite system \eqref{sir-gamma} as follows
\begin{equations}\label{sir-FP}
& \frac{\partial f_S(x,\tau)}{\partial \tau} = -K(f_S,f_I)(x,\tau) + \frac 1\e\, Q_S^\delta(f_S)(x,\tau), \\
&\frac{\partial f_I(x,\tau)}{\partial \tau} = K(f_S,f_I)(x,\tau) - \gamma f_I(x,\tau) + \frac 1\e\, Q_I^\delta(f_I)(x,\tau), \\
& \frac{\partial f_R(x,\tau)}{\partial \tau} =   \gamma f_I(x,\tau)  + \frac1\e\, Q_R^\delta(f_R)(x,\tau).
\end{equations}
The system \fer{sir-FP} is complemented with the boundary conditions \fer{bc} at $x=0$ and it contains all the information on the spreading of the epidemic in terms of the distribution of social contacts. Indeed, the knowledge of the densities $f_J(x,t)$, $J\in\{S,I,R\}$, allows to evaluate by integrations all moments of interest. However, 
in reason of the presence of the incidence rate $K(f_S,f_I)$, as given by \fer{inci}, the time evolution of the moments of the distribution functions is not explicitly computable, since the evolution of a moment of a certain order depends on the knowledge of higher order moments, thus producing a  hierarchy of equations, like in classical kinetic theory of rarefied gases \cite{Cer}. 

Before discussing the closure, \rev{i.e. how to obtain a closed set of macroscopic equations}, we establish some choices in the model in order to obtain results which are in agreement with the ones reported in \cite{Plos}. To that aim, we fix the value $\delta =1$, and the incidence rate as in \fer{simple}. In particular, the choice $\delta =1$ gives, for $J\in\{S,I,R\}$
 \be\label{q1}
 Q_J^1(f_J)(x,\tau)= \frac{\mu}{2}\frac{\partial}{\partial x}\left[\,\left(\frac x{\bar x_J} -1\right)f_J(x,\tau)\right] +\frac{\lambda}{2} \frac{\partial^2}{\partial x^2} (x f_J(x,\tau)).
 \ee
In this case \fer{para} implies $\chi = \nu$ and $\theta = \bar x_J/\nu$, and the steady  states of unit mass, for  $J\in\{S,I,R\}$, are the Gamma densities 
\be\label{gamma-e}
f_J^\infty(x;\theta, \nu) = \left(\frac \nu{\bar x_J}\right)^\nu \frac 1{\Gamma\left(\nu \right)} x^{\nu -1}
\exp\left\{ -\frac\nu{\bar x_J}\, x\right\}.
 \ee 
With this particular choice, the mean values and the \rev{energies} of the densities \fer{gamma-e}, $J\in\{S,I,R\}$, are given by
 \be\label{mv}
 \int_{\R^+} x\, f_J^\infty(x;\theta, \nu) \, dx = \bar x_J; \quad  \int_{\R^+} x^2 \, f_J^\infty(x;\theta, \nu) \, dx =\frac{\nu +1}\nu \bar x_J^2.
\ee
\rev{Following the observations of Remark \ref{rem:mean} we can now assume $\bar x_J= x_J(t)$. Also, this time dependent value can be different depending on the class to which agents belong.}
\rev{
In order to obtain the time evolution of the macroscopic observable quantities like densities and local means from the kinetic model \eqref{sir-FP}, we now consider the Fokker--Planck operator \eqref{q1}
which vanishes in correspondence to a time-dependent Gamma density equilibrium with mean $x_J(t)$.  With these notations,  system \fer{sir-FP} with $\delta =1$ reads
\begin{equations}\label{sir-FP1}
& \frac{\partial f_S(x,\tau)}{\partial \tau} = - \beta x \,f_S(x,\tau) x_I(t) \,I(\tau)  + \frac 1\e\, Q_S^1(f_S)(x,\tau), \\
&\frac{\partial f_I(x,\tau)}{\partial \tau} = \beta x\, f_S(x,\tau) x_I(t) \,I(\tau)   - \gamma f_I(x,\tau) + \frac 1\e\,  Q_I^1(f_I)(x,\tau), \\
& \frac{\partial f_R(x,t)}{\partial \tau} =   \gamma f_I(x,\tau)  + \frac1\e\, Q^1_R(f_R)(x,\tau).
\end{equations}
}
\rev{ 
Integrating both sides of equations in \fer{sir-FP1}  with respect to  $x$,  and recalling that, in presence of boundary conditions of type \fer{bc}  the Fokker-Planck type operators are mass and momentum preserving, we obtain the system for the evolution of the fractions $J$ defined in \fer{mass}, $J\in\{S,I,R\}$ 
\begin{equations}\label{sir-mass}
& \frac{d S(t)}{d t} = -\beta \, x_S(t) x_I(t) S(t)I(t),  \\
&\frac{d I(t)}{d t} = \beta\, x_S(t) x_I(t) S(t)I(t)  - \gamma I(t),  \\
& \frac{d R(t)}{d t} =   \gamma I(t),
\end{equations}
where we have restored the macroscopic time variable $ t \ge 0$.
As anticipated, unlike the classical SIR model, system \fer{sir-mass} is not closed, since the evolution of the mass fractions $J(t)$, $J \in \{S,I,R\}$,  depends on the evolution of the local mean values $x_J(t)$. 
}
The closure of system \fer{sir-mass} can be obtained by resorting, at least formally, on a limit procedure. In fact, as outlined in the Introduction, the typical time scale involved in the social contact dynamics is \rev{$\e\ll1$} which identifies a faster adaptation of individuals to social contacts with respect to the evolution time of the epidemic disease. \rev{Consequently}, the choice of the value \rev{$\e \ll1$} pushes the distribution function $f_J(x,t)$, $J\in \{S,I,R\}$ towards the Gamma equilibrium density with a mass fraction $S(t)$, \rev{respectively $I(t)$ and $R(t)$}, and \rev{local mean value $x_S(t)$, respectively $x_I(t)$ and $x_R(t)$}, as it can be easily verified from the differential expression of the interaction operators \rev{$Q_J^1$, $J \in \{S,I,R\}$}. 

Indeed, if \rev{$\e \ll 1$} is sufficiently small, one can easily argue from the exponential convergence of the solution of the Fokker-Planck equation towards the 
equilibrium $f_S^\infty(x;\theta,\nu)$ (see \cite{To4} for details), that the solution $f_S(x, t)$ remains sufficiently close to the Gamma density with mass $S(t)$ and \rev{local mean density given by $x_S(t)$ for all times}. This equilibrium distribution $f_S^\infty(x;\theta,\nu)$ can then be plugged into the first equation of \fer{sir-FP1}.
\rev{Successively, by multiplying by $x$ both sides of this equation \fer{sir-FP1} and integrating it with respect to the variable $x$, since the Fokker--Planck operator on the right-hand side is momentum-preserving, one obtains that the mean $x_S(t)S(t)$ satisfies the differential equation
\[
\frac{d }{d t} (x_S(t)S(t)) = -  \beta \, x_{S,2}(t) x_I(t) S(t)I(t), 
\]
which depends now on the second order moment. \rev{However,} it is now possible to close this expression by using the energy of the local equilibrium distribution, which can be expressed in terms of the mean value as in \fer{mv} as follows
\[
 x_{S,2}(t) = \frac{\nu+1}{\nu} x_S^2(t).
\]
Therefore, we have
 \[
S(t) \frac{d x_S(t)}{ d t} = - \beta \, x_{S,2}(t) x_I(t) S(t)I(t) - x_S(t)\frac{dS(t)}{d t},
 \]
where the time evolution of the fraction $S(t)$ can be recovered by the first equation of system \fer{sir-mass}. Hence, the evolution of the local mean value $x_S(t)$ satisfies the equation
\be\label{okkk}
\frac{d x_S(t)}{d t} = - \frac\beta\nu x_{S}(t)^2 x_I(t) I(t).
\ee
 An analogous  procedure can be done with the second equation in system \fer{sir-FP1}, which leads to relaxation towards a Gamma density with mass fraction $I(t)$ and local mean value given by $x_I(t)$, and with the third equation in system \fer{sir-FP1}.
We easily obtain in this way the system that governs the evolution of the local mean values of the social contacts of the classes of susceptible, infected and recovered individuals
 \begin{equations}\label{sir-mean}
& \frac{d x_S(t)}{d t} = - \frac\beta\nu x_{S}(t)^2 x_I(t) I(t),  \\
&\frac{d x_I(t)}{d t} =   \beta x_{S}(t) x_I(t)  \left( \frac{1+\nu}{\nu} x_S(t) - x_I(t) \right) S(t),  \\
& \frac{d x_R(t)}{d t} =  \gamma \frac{I(t)}{R(t)}\left( x_I(t) - x_R(t)\right).
\end{equations}
 The closure of the kinetic system \fer{sir-gamma} around a Gamma-type equilibrium of social contacts leads then to a system of six equations for the pairs of mass fractions $J(t)$ and local mean values $x_J(t)$, $J\in\{S,I,R\}$. In the following, we refer to the coupled systems \fer{sir-mass} and \fer{sir-mean} as the social SIR model (S-SIR).} 
\rev{
With respect to the classical epidemiological model from which we took inspiration, the main novelty is represented by the presence of system \fer{sir-mean}, that describes the evolution of the social contacts. It is immediate to conclude from the first equation of \fer{sir-mean} that the local mean number of contacts of the population of susceptible individuals decreases, thus showing that the \textit{social answer} to the presence of the pandemic is to reduce the number of social contacts. A maybe unexpected behavior is present in the second equation of \fer{sir-mean}, which indicates that, at least in the initial part of the time evolution of the S-SIR model, the class of infected individuals increases the local mean number of social contacts. This effect must be read as a consequence of the more probable transition from susceptible to infected of individuals with a high number of social contacts.
}

\rev{
It is interesting to remark that system \fer{sir-mean} is explicitly dependent on the positive parameter $\nu =\lambda/ \sigma$, which measures the heterogeneity of the population in terms of the variance of the statistical distribution of social contacts. More precisely, small values of the constant $\nu$ correspond to high values of the variance, and thus to a larger heterogeneity of the individuals with respect to social contacts. This is an important point which is widely present and studied in the epidemiological literature \cite{AM, Diek, DH}. Concerning the COVID-19 pandemic, the influence of population heterogeneity on herd immunity has been recently quantified in \cite{BBT} by testing a SEIR model on different types of populations categorized by different ages and/or different activity levels, thus expressing different levels of heterogeneity. }

\rev{
A limiting case of system \fer{sir-mean} is obtained by letting the parameter $\nu \to +\infty$, which corresponds to push the variance to zero (absence of heterogeneity). In this case, if the whole population starts with a common number of daily contacts, say $\bar x$, it is immediate to show that the number of contacts remains fixed in time, thus reducing system \fer{sir-mass} to a classical SIR model with contact rate $\beta \bar x^2$. Hence this classical epidemiological model is contained in \fer{sir-mass}-\fer{sir-mean} and corresponds to consider the non realistic case of a population that, regardless of the presence of the epidemic, maintains the same fixed number of daily contacts. The described behaviors are exemplified in Figure \ref{fig:system} where we considered $S(0) = 0.98$, $I(0) = R(0) = 0.01$ and mean number of contacts $x_S(0) = x_I(0) = x_R(0) = 15$ for two choices $\nu = 0.5$ and $\nu = 1$. We can easily observe how the number of recovered is affected by contact heterogeneity: the smaller the heterogeneity is, the larger the population recovers from the pandemic. 
\begin{figure}
\centering
\includegraphics[scale = 0.4]{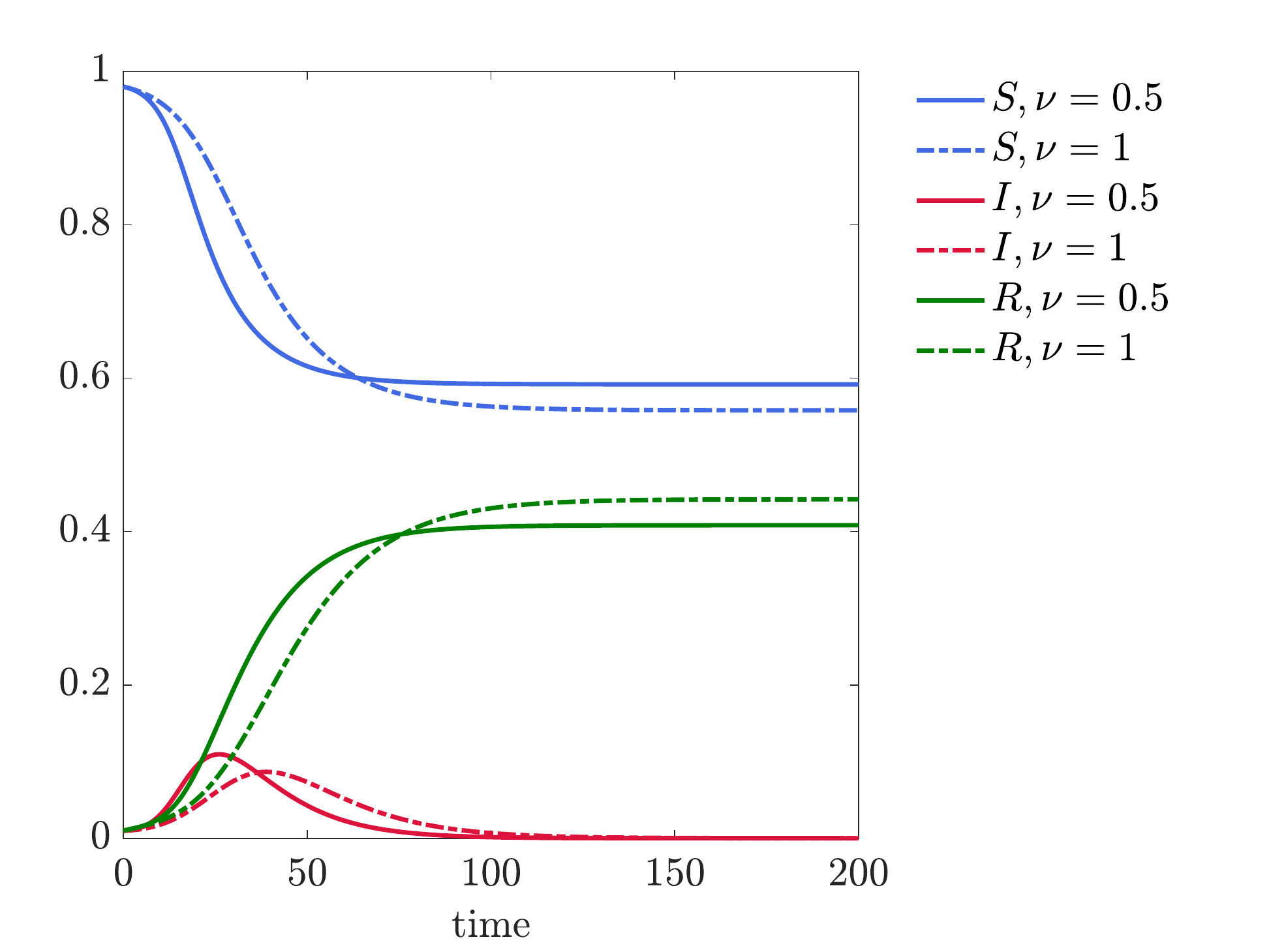}
\includegraphics[scale = 0.4]{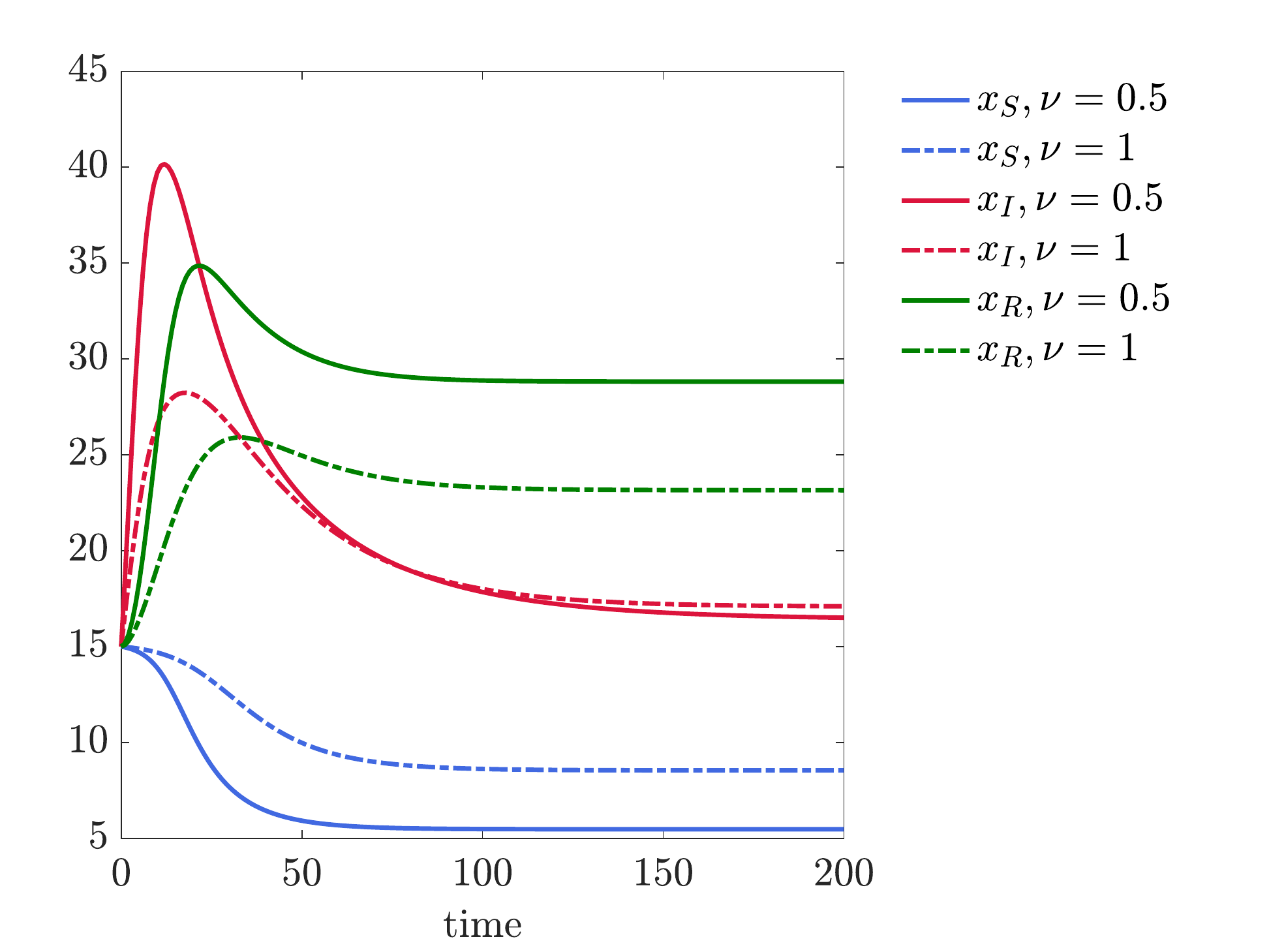}
\caption{Evolution of system \fer{sir-mass}-\fer{sir-mean} for $\nu = 0.5$ and $\nu = 1$. }
\label{fig:system}
\end{figure}
\begin{remark} The derivation leading to systems \fer{sir-mass} and \fer{sir-mean} can be easily generalized to local incidence rates \fer{inci} with a contact function of the form $\kappa(x,y) = \beta x^\alpha y^\alpha $ with $\alpha \not= 1$. In this case, however, we do not have an explicit formula which connects the moment of order $\alpha$ of the Gamma density with the moment of order $\alpha +1$. Also, the procedure can be applied to  equilibria which are different from the Gamma density considered here, provided this density has enough moments bounded. 
\end{remark}
\begin{remark} The approach just described can be easily adapted to other compartmental  models in epidemiology like SEIR, MSEIR \cite{BCF, DH, HWH00} and/or SIDHARTE \cite{Gatto,Bruno}. For all these cited models, the fundamental aspects of the interaction between social contacts and the spread of the infectious disease we expect not to change in a substantial way. 
\end{remark} 
}
\subsection{\rev{A social-SIR model with saturated incidence rate} }

 \rev{The system \fer{sir-mass}-\fer{sir-mean} is a model for describing the time evolution of an epidemic in terms of the statistical distribution of social contacts, without taking into account any external intervention. However, protection measures such as lockdown strategies inevitably cause reduction in the average number of social contacts of the population which can be taken explicitly into account in our model. In the epidemiological literature, a natural way to introduce this mechanism, which dates from the work of Capasso and Serio \cite{Capasso}, consists in considering a non linear incidence rate whose main feature is to be bounded with respect to the number of infected. 
 Interestingly, on this subject, we aim to highlight that a similar behavior of the incidence rate can be directly derived starting from our social-SIR model \fer{sir-mass}-\fer{sir-mean}. The additional hypothesis that is sufficient to introduce is that 
 	the average number $\tilde x_I$ of social contacts of infected is frozen as an effect, for instance, of external interventions aimed in controlling the pandemic spread. If this is the case, one can explicitly solve the first equation of system \fer{sir-mean}, which now reads
 \be\label{freeze}
 \frac{d x_S(t)}{d t} = - \frac\beta\nu x_{S}(t)^2 \tilde x_I I(t), 
\ee 
due to the fact that $x_I(t) = x_I(t=0)= \tilde x_I$. 
The exact solution of the equation \fer{freeze} can then be computed and it gives
 \be\label{sol-freeze}
 x_S(t) = \frac{x_S(0)}{1 + \displaystyle\frac \beta\nu x_S(0) \int_0^t I(s)\, ds}.
 \ee
 The above expression is a generalization of the so-called saturated incidence rate\cite{Capasso, KM} whose \textit{classical} form is 
 \be\label{satincidence}
 g(I)=\frac{1}{1+\phi I}
 \ee 
 with $\phi$ a suitable positive constant. The same expression can be found from \eqref{sir-mass} by plugging \eqref{sol-freeze} into the system. This gives
 \begin{equations}\label{sir-mass-closed}
 & \frac{d S(t)}{d t} = -\bar\beta \, S(t)I(t) H(I(t),t),  \\
 &\frac{d I(t)}{d t} = \bar\beta\,  H(I(t),t) S(t)I(t)  - \gamma I(t),  \\
 & \frac{d R(t)}{d t} =   \gamma I(t),
\end{equations}
where $\bar\beta =\beta x_S(0)\tilde x_I$ and 
\be\label{ese1}
H(r(t),t) = \frac{1}{1 + \displaystyle \frac{\bar\beta}{\nu\tilde x_I} \int_0^t r(s)\, ds}, \quad 0<r \le 1.
\ee
Finally by approximating the integral $\int_0^t r(s)\, ds\approx t r(t)$ one obtains 
\be\label{ese2}
H(r(t),t) = \frac{1}{1 + \displaystyle \phi(t) r(t)}, \quad 0<r \le 1.
\ee
with $\phi(t)=(\bar\beta t)/(\nu\tilde x_I)$. In the next Section \ref{numerics} we will perform some numerical results in which the social SIR model \eqref{sir-mass-closed} is used both in the case in which the incidence rate takes the form \eqref{ese1} as well as in the classical case \ref{ese2} with $\phi(t)=\phi$. 
Let us now note that by defining
\be\label{new-inci}
 D(S(t),I(t)) = \int_{\R^+} K(f_S,f_I)(x, t) \, dx = \bar\beta  H(I(t),t)S(t)I(t).
 \ee 
we have that in both cases $D(S,I)$ fulfils all the properties required by the class of non-linear incidence rates considered in  \cite{KM}. Indeed, $D(S,0) = 0$, and the function $D(S,I)$ satisfies 
 \be\label{pro1}
  \frac{\partial D(S,I)}{\partial I}>0, \quad  \frac{\partial D(S,I)}{\partial S} >0
 \ee 
for all $S,I >0$. Moreover  $D(S,I)$ is concave with respect to the variable $I$, i.e.
 \be\label{pro2}
 \frac{\partial^2D(S,I)}{\partial I^2} \le 0,
 \ee
for all $S,I >0$. 
}


\begin{figure}\centering
	{\includegraphics[scale = 0.5]{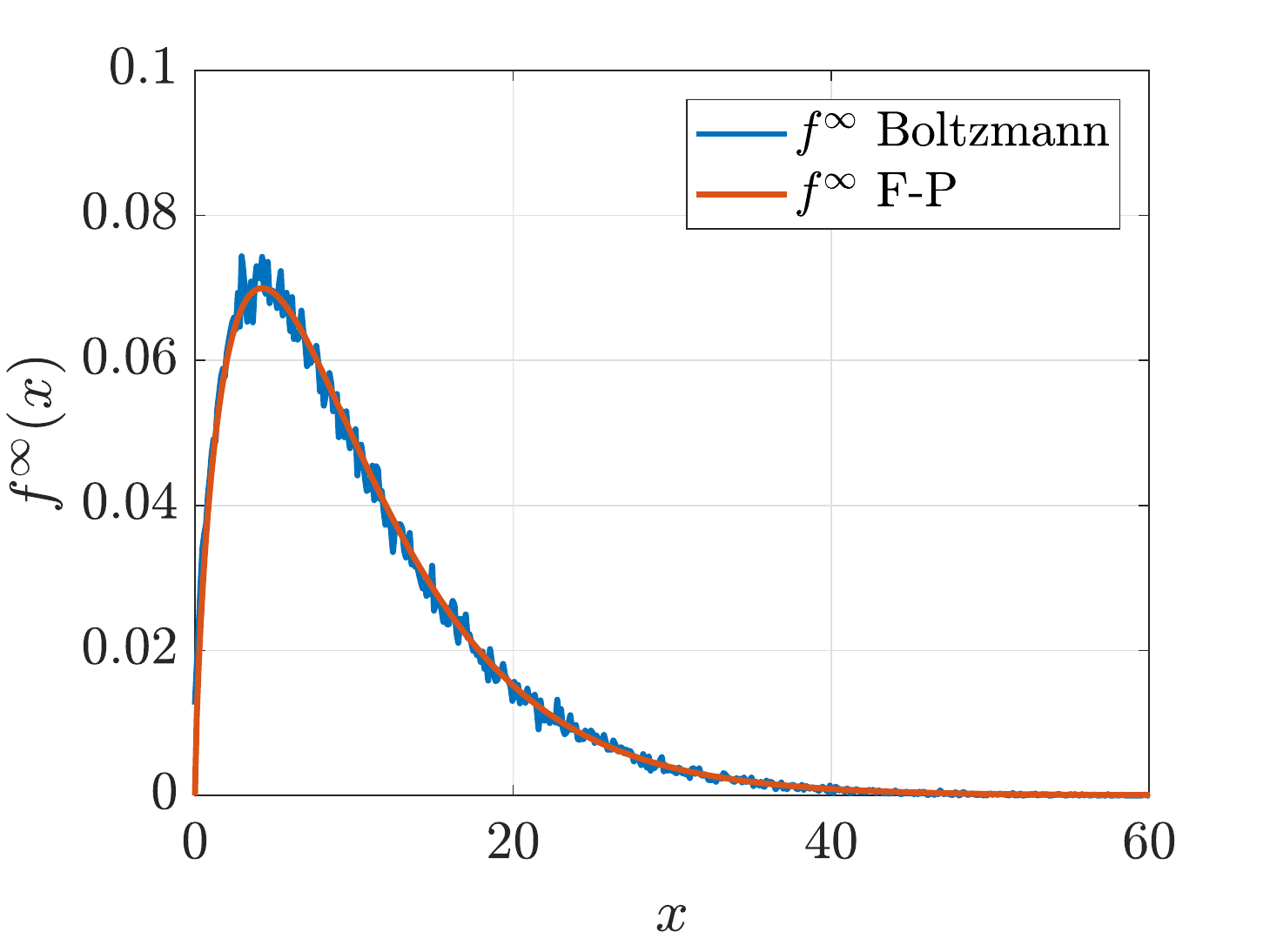}}
	\caption{\textbf{Test 1}. Large time distribution of the Boltzmann dynamics compared with the equilibrium state of the corresponding Fokker-Planck equation. The initial distribution has been chosen of the form in \eqref{eq:f0_num}.  }\label{fig:test1}
\end{figure}

%
%

\section{Numerical experiments}\label{numerics}

\begin{figure}
\centering
	{   \includegraphics[scale = 0.4]{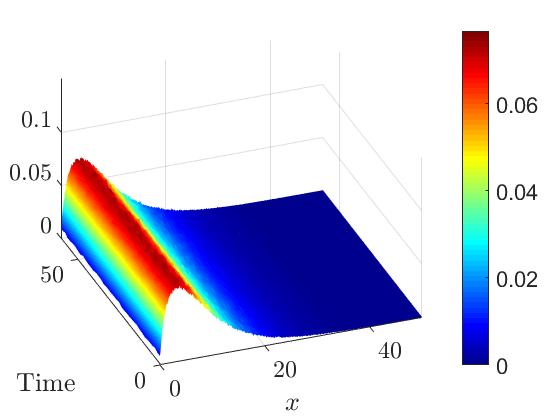}
		\includegraphics[scale = 0.4]{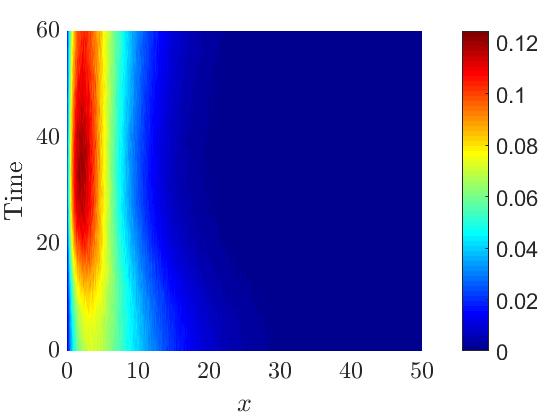}\\
		\includegraphics[scale = 0.5]{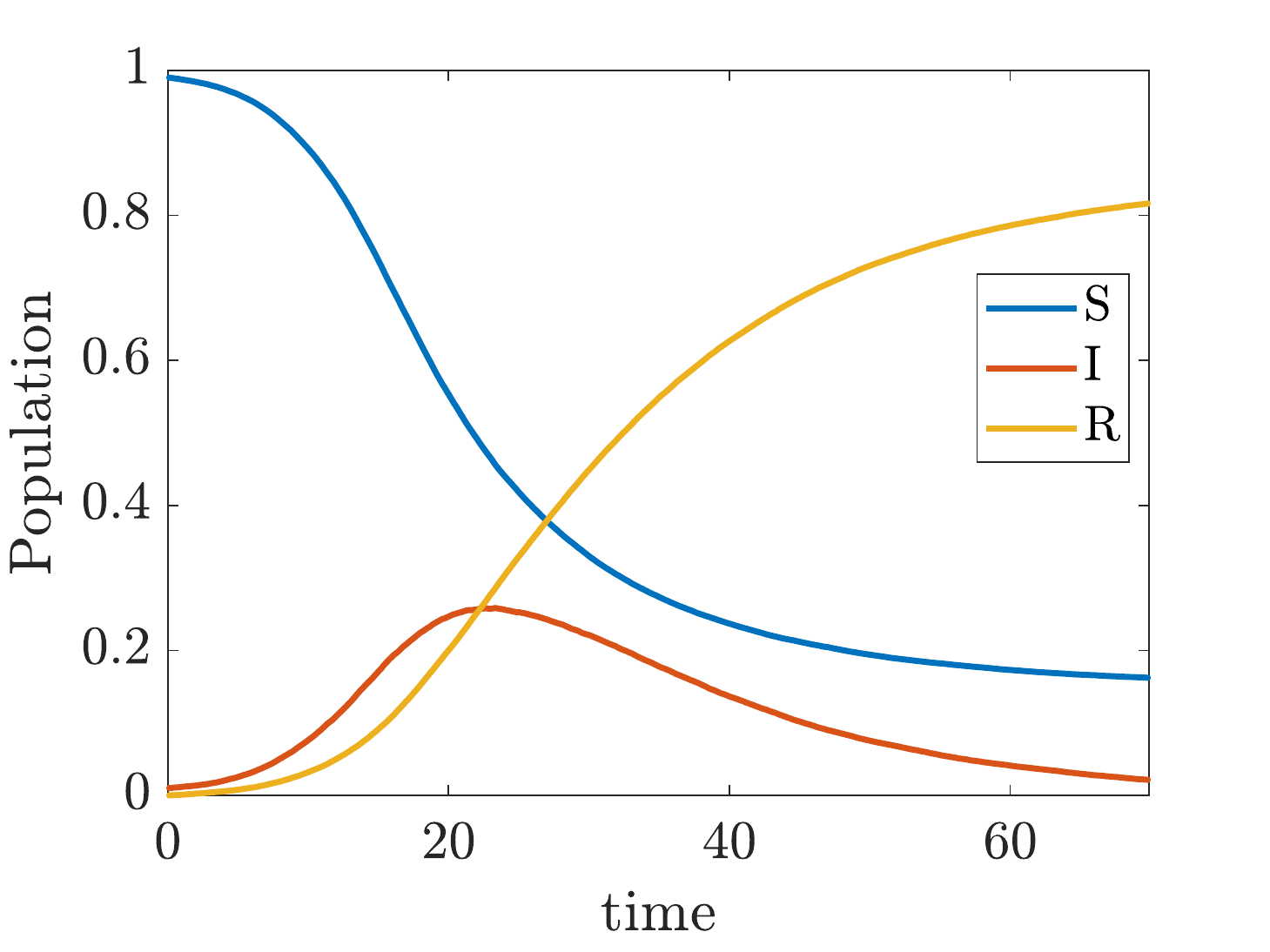}        
		\includegraphics[scale = 0.5]{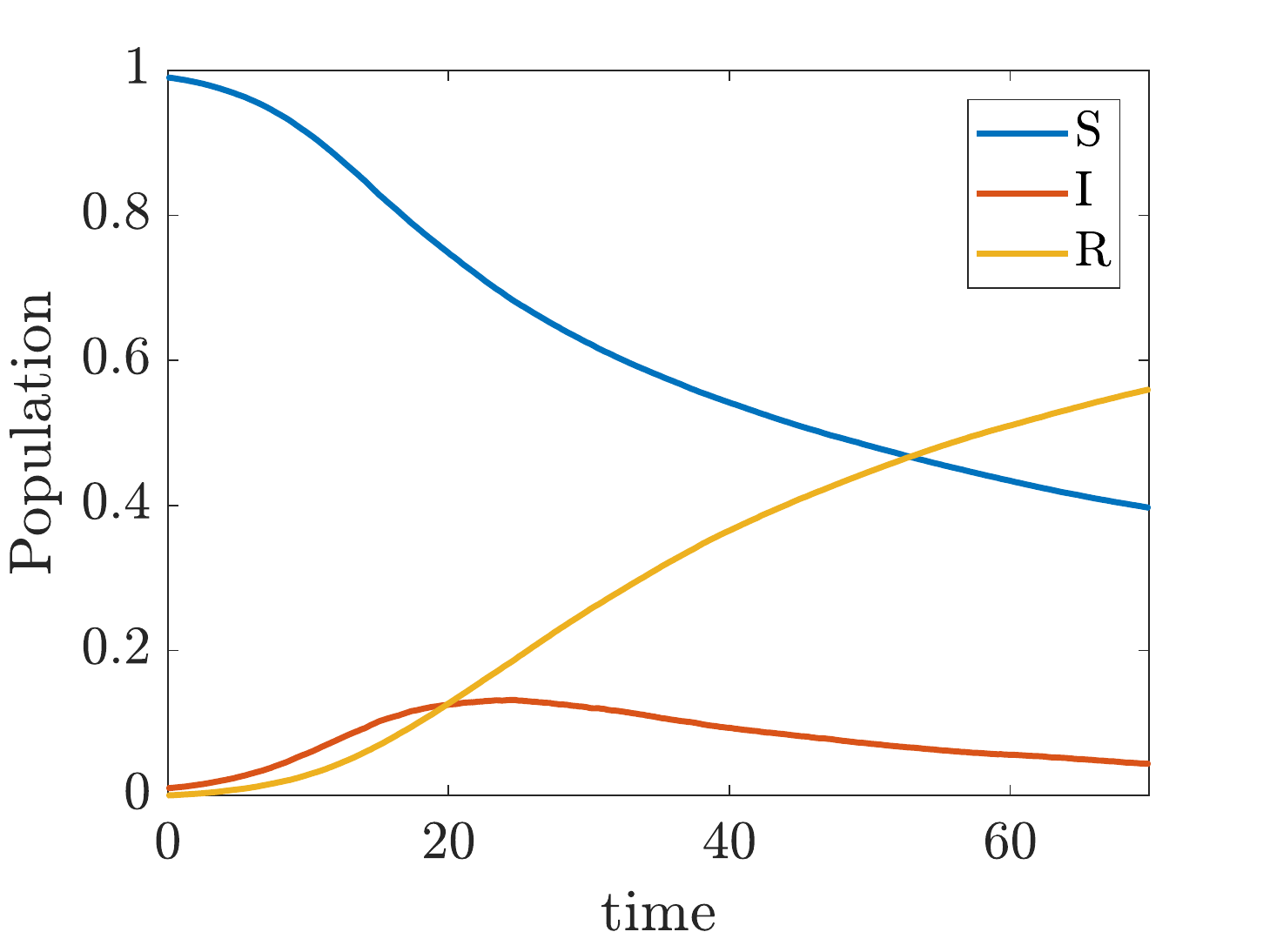}\\
		\includegraphics[scale = 0.5]{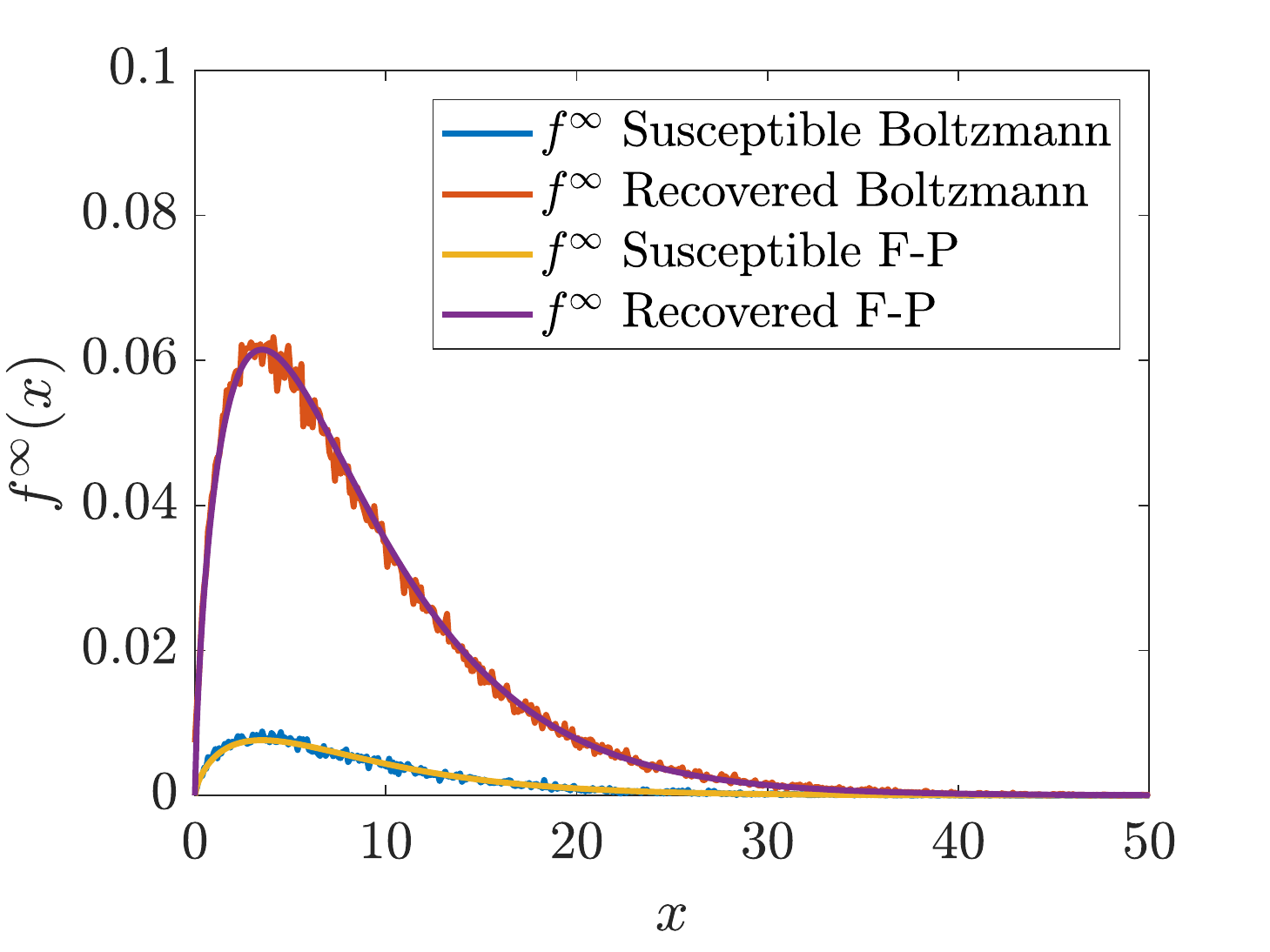}
		\includegraphics[scale = 0.5]{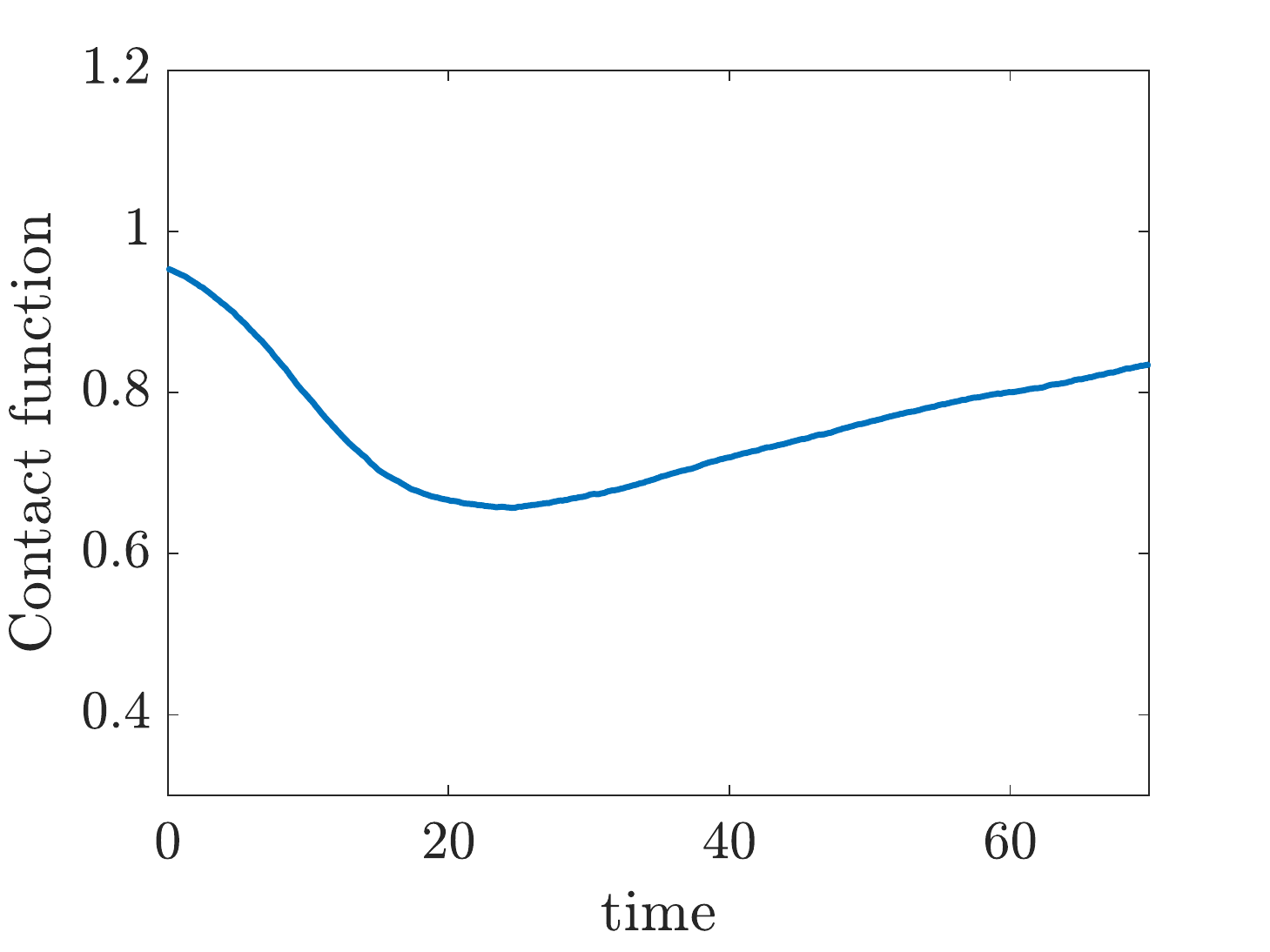}
	}
	\caption{\textbf{Test 1}. Top: distribution of the daily social contacts for the two choices of the function $H$. Middle: SIR dynamics corresponding to the different choices of the different mean number of daily contact (left constant case, right as a function of the epidemic). Bottom left: final distribution of the number of contacts. Bottom right: time evolution of the contact function $H$.}\label{fig:test1_2}
\end{figure}

In addition to analytic expressions, numerical experiments allow us to visualize and quantify the effects of social contacts on the SIR dynamics used to describe the time evolution of the epidemic. More precisely, starting from a given equilibrium distribution detailing in a probability setting the daily number of contacts of the population, we show how the coupling between social behaviors and number of infected people may modify the epidemic by slowing down the number of encounters leading to infection. In a second part, we discuss how some external forcing, mimicking political choices and acting on restrictions on the mobility, may additionally improve the reduction of the epidemic trend avoiding concentration in time of people affected by the disease and then decreasing the probability of hospitalization peaks. In a third part, we focus on experimental data for the specific case of COVID-19 in different European countries and we extrapolate from them the main features characterizing the contact function $H(\cdot)$. 

\subsection{Test 1: On the effects of the social contacts on the epidemic dynamics}\label{numericsI}
We solve the social-SIR kinetic system \eqref{sir-FP1}. The starting point is represented by a population composed of $99.99\%$ of susceptibles and $0.01\%$ of infected. The distribution of the number of contact is described by \eqref{equilibrio} with $\nu=1.65$, $\delta=1$ and $x_J=10.25$ in agreement with \cite{Plos} while the epidemic parameters are $\beta=0.25/x_J^2$ and $\gamma=0.1$. The kinetic model \eqref{sir-FP} is solved by a splitting strategy together with a Monte Carlo approach where the number of samples used to described the population is fixed to $M=10^6$. The time step is fixed to $\Delta t=10^{-2}$ and the scaling parameter is $\tau=10^{-2}$. These choices are enough to observe the convergence of the Boltzmann dynamics to the Fokker-Planck one as shown in Figure~\ref{fig:test1} where the analytical equilibrium distribution is plotted together with the results of the Boltzmann dynamics. We considered also uniform initial distribution 

\be\label{eq:f0_num}
f_0(x) = \dfrac{1}{c} \chi(x \in [0,c]), \quad c=20,
\ee
being $\chi(\cdot)$ the indicator function. In the introduced setting, we then compare two distinct cases: in the first one we suppose that \rev{nonlinear effects in the contacts dynamics do not modify the contact rate,}
meaning $H(I(t),t)=1$, while the second includes the effects of the function $H(I(t),t)$ given in \eqref{ese2} with $\alpha=10$. The results are depicted in Figure~\ref{fig:test1_2}. The top right images show the time evolution of the distribution of the number of contacts for the two distinct cases, while the middle images report the corresponding evolution of the epidemic. For this  the second case, the function $H(I(t),t)$ as well as the distribution of contacts for respectively the susceptible and the recovered are shown at the bottom of the same figure. We clearly observe a reduction of the peak of infected in the case in which the dynamics depends on the number of contacts with $H(I(t),t)$ given by \eqref{ese2}. We also observe a spread of the number of infected over time when sociality reduction is taken into account.

\subsection{Test 2: Forcing a change in the social attitudes.}\label{numericsII}

Next, we compare the effects on the spread of the disease when the population adapts its habits with a time delay with respect to the onset of the epidemic. This kind of dynamics corresponds to a modeling of a possible lockdown strategy whose effects are to reduce the mobility of the population and, correspondingly, to reduce the number of daily contacts in the population.

The setting is similar to the one introduced in Section~\ref{numericsI} and we consider a \rev{switch between $H=1$ to $H(I(t))=1/(1+\alpha I(t))$ when the number of infected increases}. The social parameters are $\nu=1.65$, $\delta=1$ and $x_J=10.25$, as before, while the epidemic parameters are $\beta=0.25/x_J^2$ and $\gamma=0.1$, the final time is fixed to $T=70$. The initial distribution of contacts is also assumed to be of the form \eqref{eq:f0_num}.

\begin{figure}
\centering
	{\includegraphics[scale = 0.5]{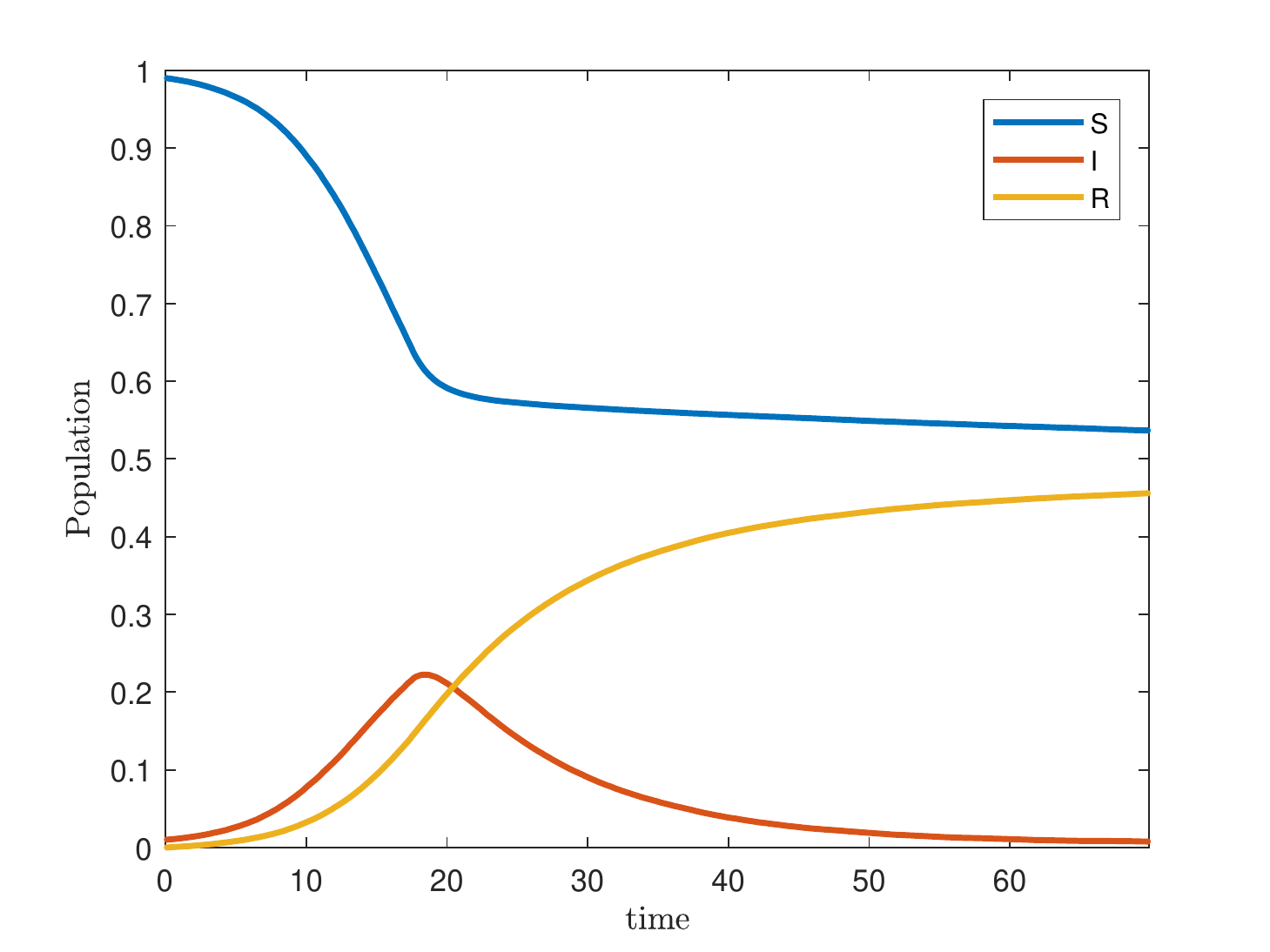}
		\includegraphics[scale = 0.45]{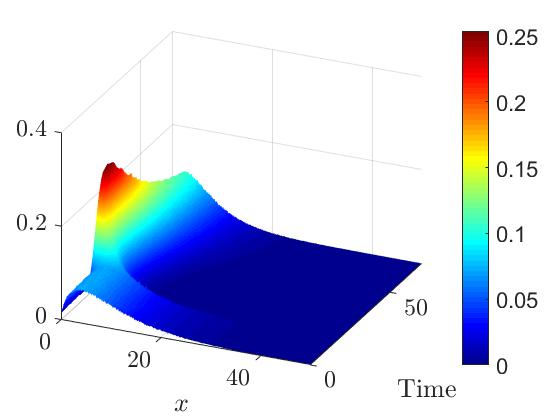}\\
		\includegraphics[scale = 0.5]{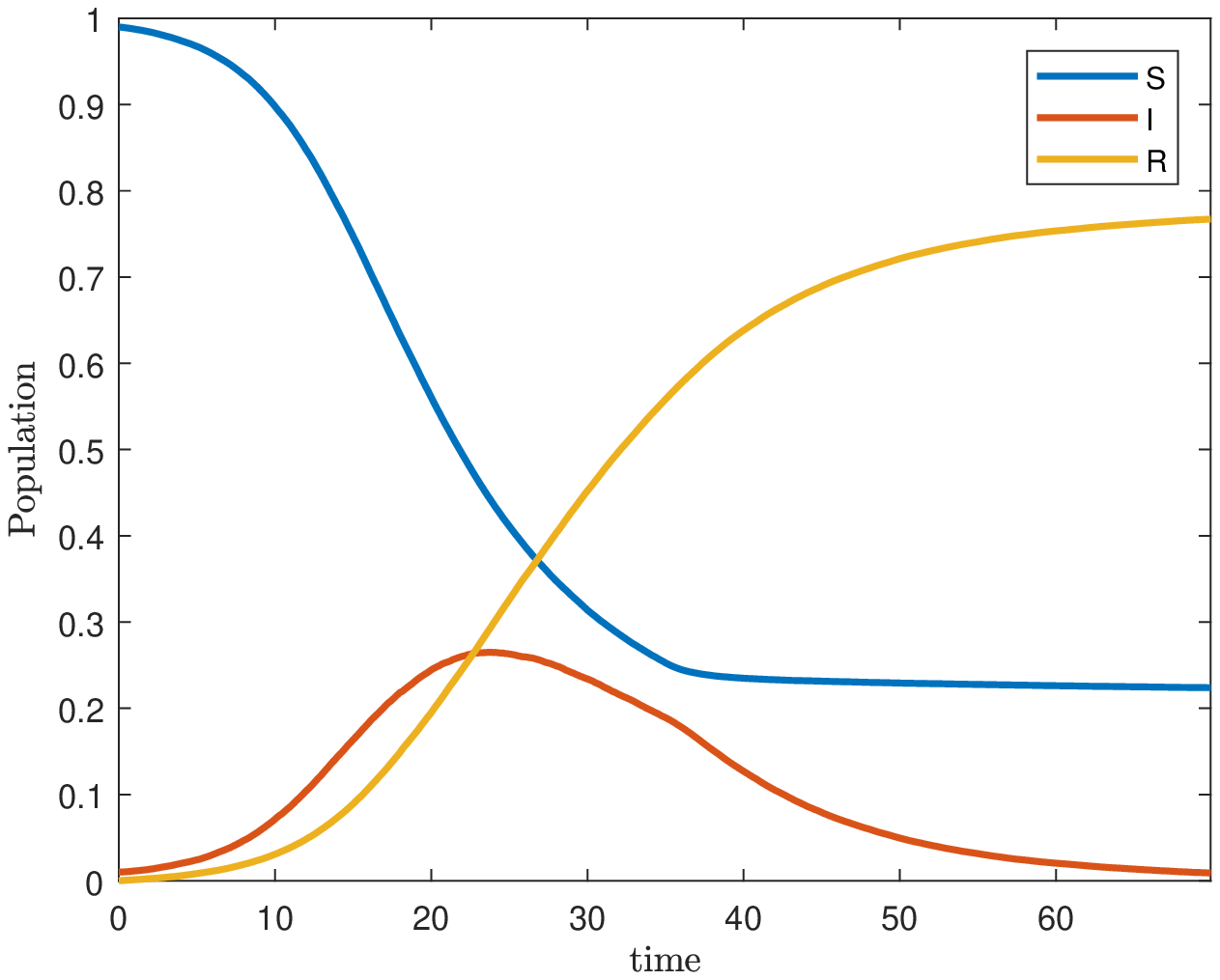}
		\includegraphics[scale = 0.45]{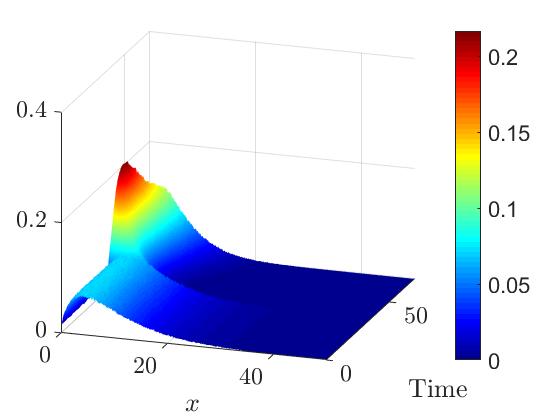}\\
		\includegraphics[scale = 0.5]{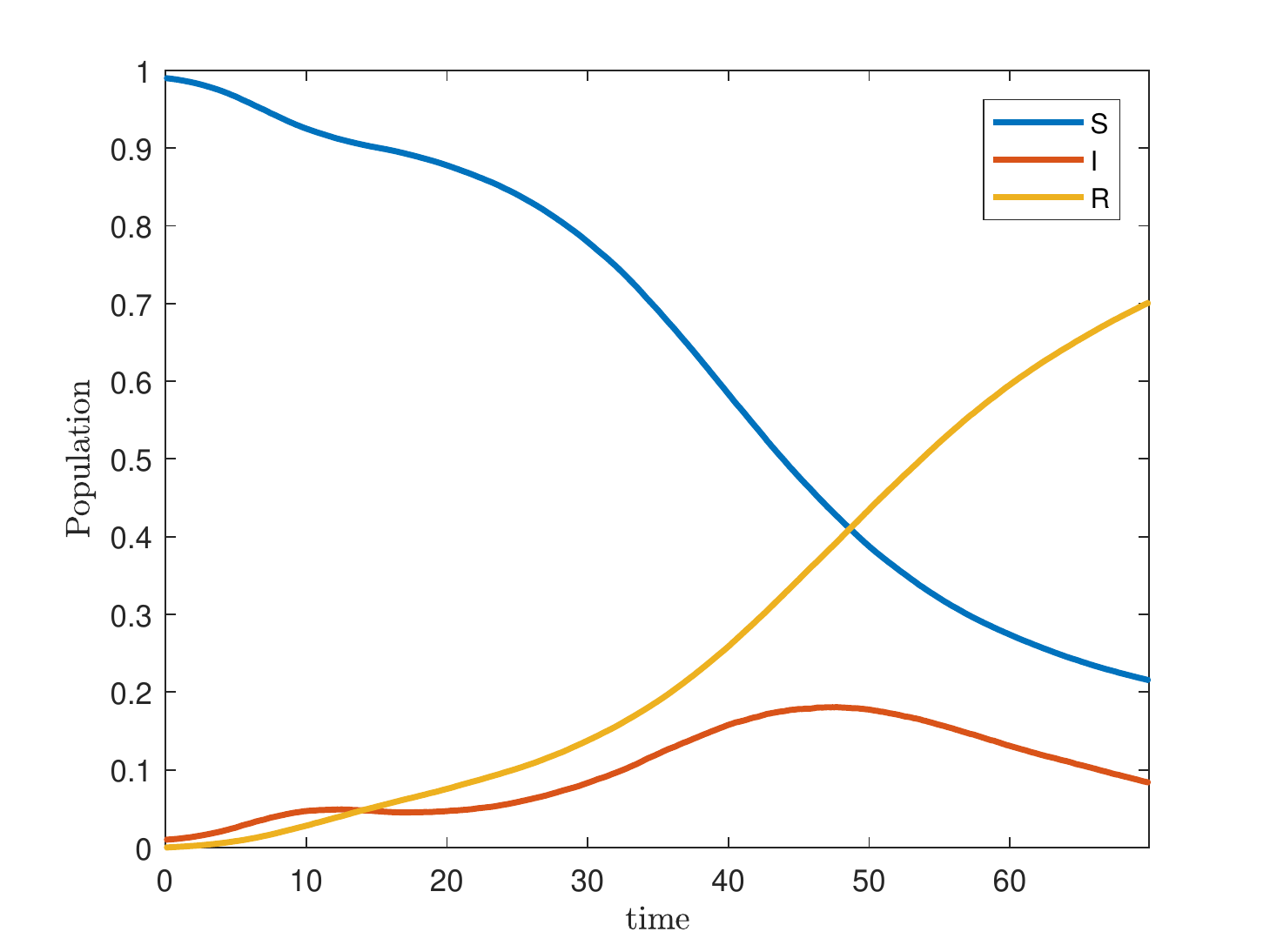}
		\includegraphics[scale = 0.455]{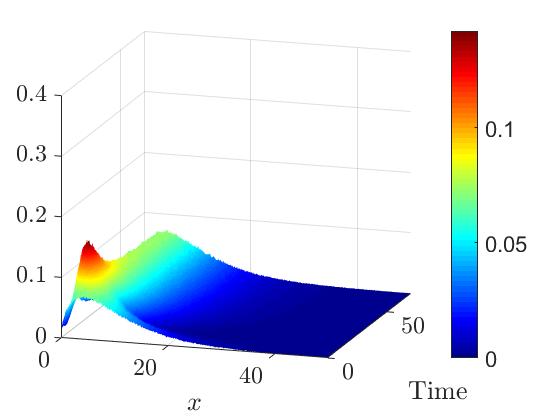}
	}
	\caption{\textbf{Test 2}. Comparisons of different lockdown behaviors. Top: late lockdown. Middle: early lockdown. Bottom: early lockdown and successive relaxation.}\label{fig:test2}
\end{figure}

We consider three different settings, in the first one $H=1$ up to $t<T/4$, in the second one up to $t<T/2$ while in the third one we prescribe a lockdown for a limited amount of time ($5<t<15$) and then we relax back to $H=1$. The results are shown in Figure~\ref{fig:test2} for both the distribution of daily contacts over time and the SIR evolution. We can identify three scenarios. The first on the top causes a clear change to the epidemic dynamic, an inversion around $t=20$ happens. For the second, we observe a reduction of the speed of the infection, while for the third case we first observe inversion and then the resurgence of the number of infected when the lockdown measures are relaxed.

\subsection{Test 3: Extrapolation of the contact function from data}\label{numericsIII}
In this part, we consider data about the dynamics of COVID-19 in three European countries: France, Italy and Spain. For these three countries, the evolution of the disease, in terms of reported cases, evolved in rather different ways. The estimation of epidemiological parameters of compartmental models is an inverse problem of generally difficult solution for which different approaches can be considered. We mention in this direction a very recent comparison study \cite{Liu}. It is also worth to mention that often the data are partial and heterogeneous with respect to their assimilation, see for instance discussions in \cite{APZ,Cetal,Chowell,Rob}. This makes the fitting problem  challenging and the results naturally affected by uncertainty.

The data concerning the actual number of infected, recovered and deaths of COVID-19 are publicly available from the John Hopkins University GitHub repository \cite{DDG}. For the specific case of Italy, we considered instead the GitHub repository of the Italian Civil Protection Department \cite{Prot}. In the following, we present the adopted approach which is based on a strategy with two optimisation horizons (pre-lockdown and lockdown time spans) depending on the different strategies enacted by the governments of the considered European countries.

\begin{figure}
	\centering
	\includegraphics[scale = 0.28]{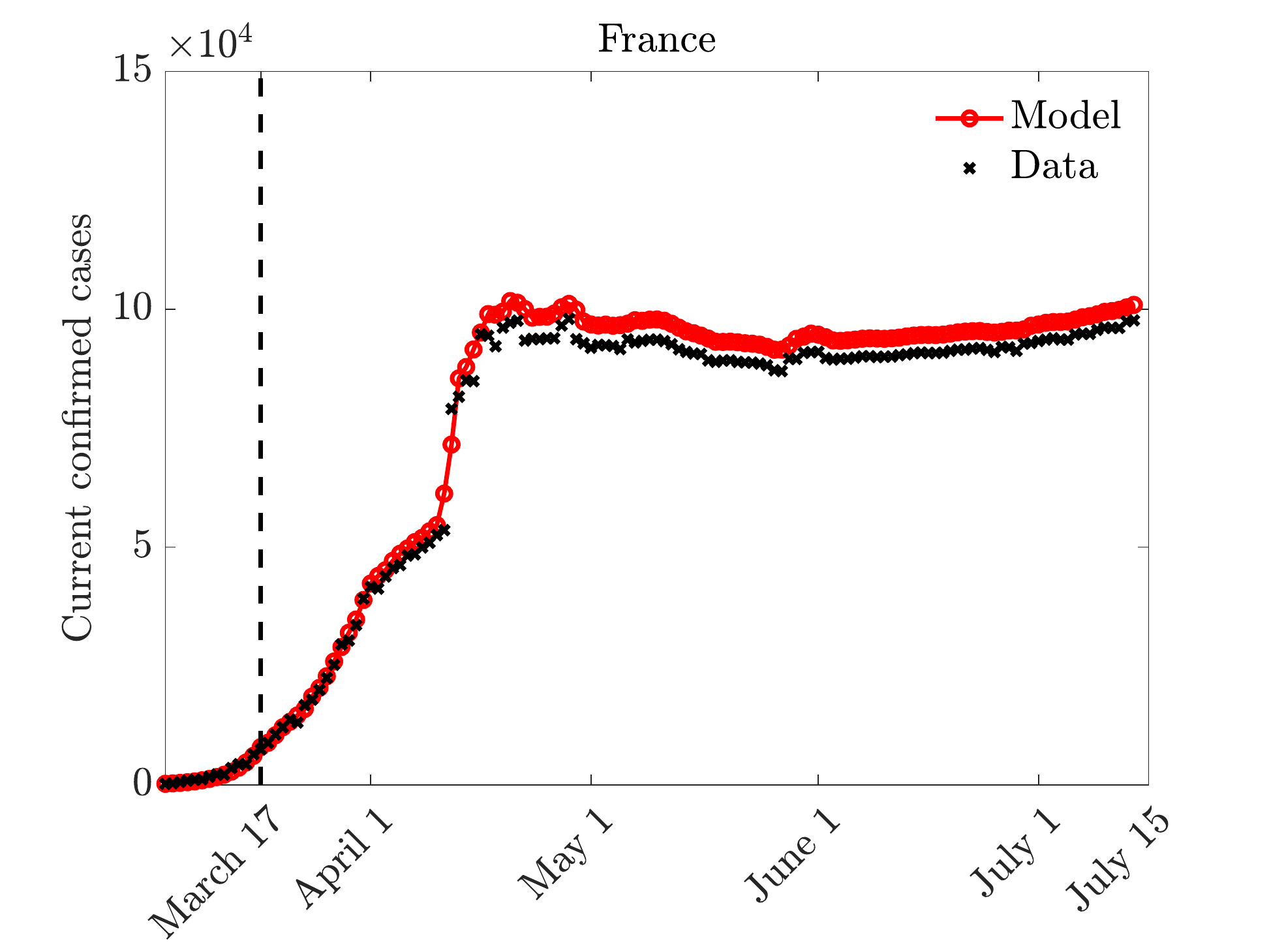}
	\includegraphics[scale = 0.28]{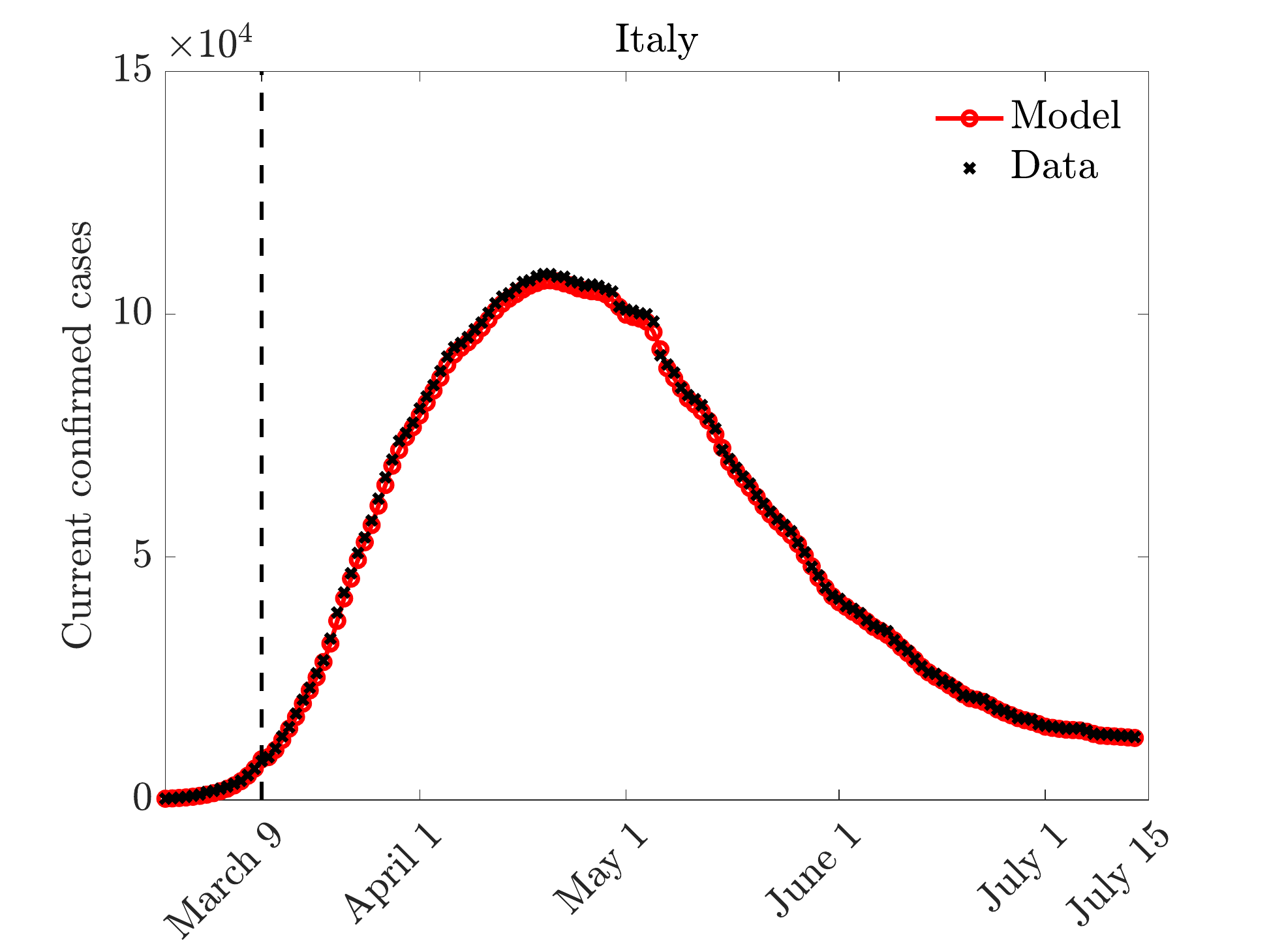}
	\includegraphics[scale = 0.28]{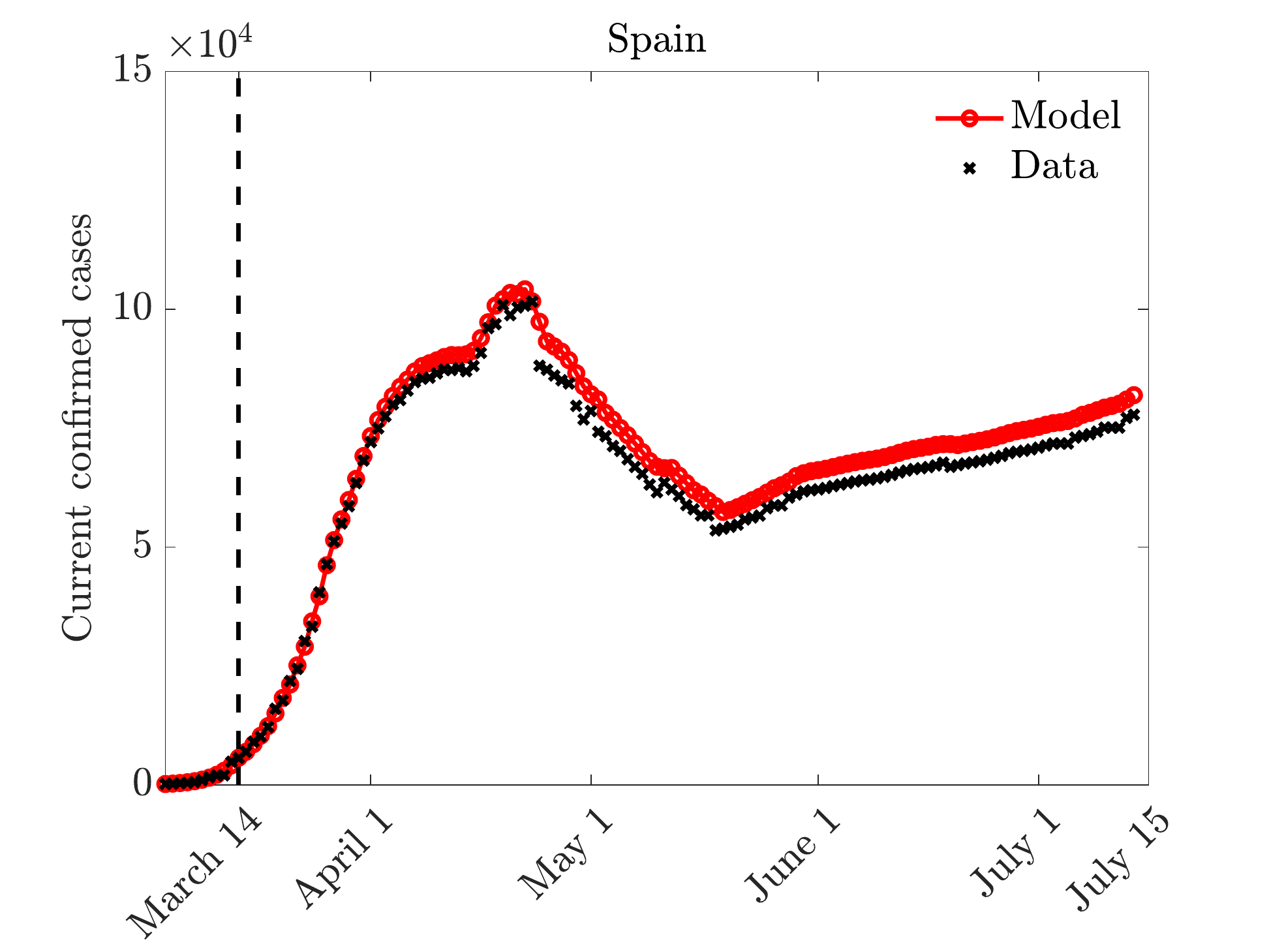}
	\caption{\textbf{Test 3}. Fitting of the parameters of model \eqref{sir-mass-closed} where $\beta,\gamma>0$ were estimated before the lockdown measures assuming $H \equiv 1$. The parameters characterizing the function $H(\cdot)$ in \eqref{sir-mass-closed} have been computed during and after lockdown at regular interval of time, up to July 15. The lockdown measures change in each county (dashed line). }
	\label{fig:t3_1}
\end{figure}

In details, we considered first the time interval $t \in [t_0,t_\ell]$, being $t_\ell$ the day in which lockdown started in each country (Spain, Italy and France) and $t_0$ the day in which the reported cases hit 200 units. The lower bound $t_0$ has been imposed to reduce the effects of fluctuations caused by the way in which data are measured which have a stronger impact when the number of infected is low. Once the time span has been fixed, we then considered a least square problem based on the minimization of a cost functional $\mathcal J$ which takes into account the relative $L^2$ norm of the difference between the reported number of infected and the reported total cases $\hat I(t)$, $\hat I(t)+\hat R(t)$ and the evolution of $I(t)$ and $I(t)+R(t)$ prescribed by system \eqref{sir-mass-closed} with $H \equiv 1$. In practice, we solved the following constrained optimisation problem 
\begin{equation}\label{eq:min_probl}
\textrm{min}_{\beta,\gamma} \mathcal J(\hat I, \hat R, I, R)
\end{equation}
where the cost functional $\mathcal J$ is a convex combination of the mentioned norms and assumes the following form 
\[
J(\hat I, \hat R, I, R) = p \frac{ \| \hat I(t)-I(t)\|_2}{\| \hat I(t) \|_2}+ (1-p) \frac{ \| \hat I(t)+\hat R(t)-I(t)-R(t) \|_2}{\| \hat I(t) + \hat R(t) \|_2}
\]

\begin{table}[t]
	\begin{center}
		\begin{tabular}{c | c | c |  c }
			& $\underset{\textrm{Mar 5-Mar 17}}{\textrm{France}}$& $\underset{\textrm{Feb 24-Mar 9}}{\textrm{Italy}}$ & $\underset{\textrm{Mar 5-Mar 14}}{\textrm{Spain}}$  \\ 
			\hline\hline
			& & & \\[-.3cm]
			$\beta$  & $0.300686$ & $0.317627$& $0.370362$ \\
			\hline
			$\gamma$ & $0.040000$ & $0.058391$ & $0.043168$ \\
			\hline
			$R_0$ & $7.5172$ & $5.4397$ & $8.5795$ 
		\end{tabular}
	\end{center}
	\caption{\textbf{Test 3}. Model fitting parameters in estimating \rev{the reproduction number} for the COVID-19 outbreak before lockdown in various European countries.}
	\label{tab:kcont}
\end{table}

We choose $p = 0.1$ and we look for a minimum under the constraints
\[
\begin{split}
0\le \beta \le 0.6, \qquad 0.04 \le \gamma \le 0.06.
\end{split}
\]
In Table \ref{tab:kcont} we report the results of the performed parameter estimation together with the resulting \rev{reproduction number} $R_0$ defined in \eqref{r0}. 

Once that the contagion parameters have been estimated in the pre-lockdown time span, we successively proceeded with the estimation of the shape of the function $H$ from the data.
To estimate this latter quantity, we solved the following optimization problem
\begin{equation}\label{eq:min_probl2}
\textrm{min}_H \mathcal J
\end{equation}
in terms of $H$
where $\mathcal J$ is the same functional of the previous step and 
where in the evolution of the macroscopic model the values $\beta,\gamma$ have been fixed as a result of the first optimization in the pre-lockdown period. The parameters chosen for \eqref{eq:min_probl2} are $p = 0.1$ while the constraint is
\[
\begin{split}
0\le H \le 1. 
\end{split}
\]
The second optimisation problem has been solved up to last available data for each country with daily time stepping $h = 1$ and over a time window of three days. This has been done with the scope of regularizing possible errors due to late reported infected and smoothing the shape of $H$. Both optimisation problems \eqref{eq:min_probl}-\eqref{eq:min_probl2} have been tackled using the Matlab functions \texttt{fmincon} in combination with a RK4 integration method of the system of ODEs.

\begin{figure}
	\centering
	\includegraphics[scale = 0.4]{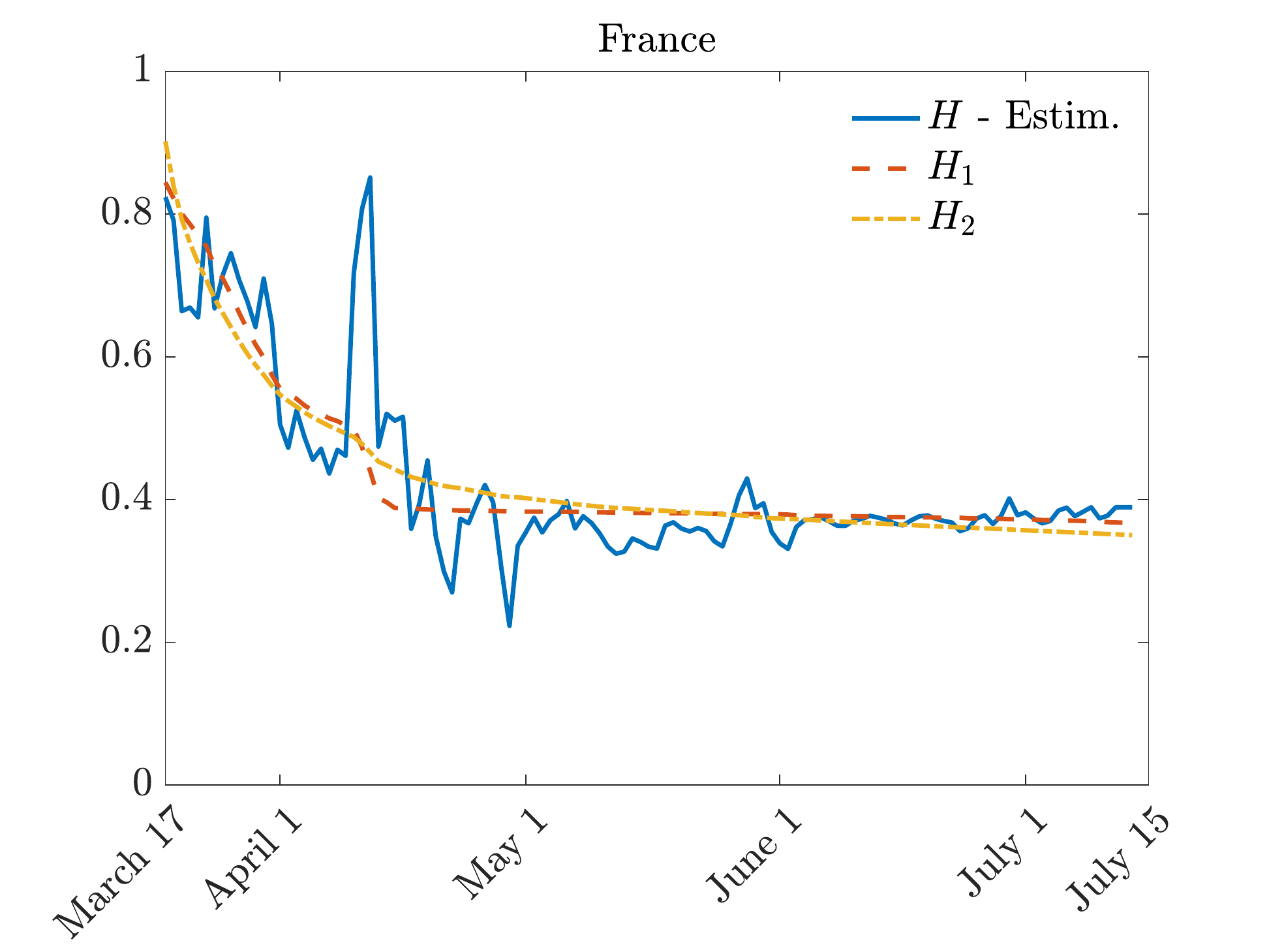}
	\includegraphics[scale = 0.4]{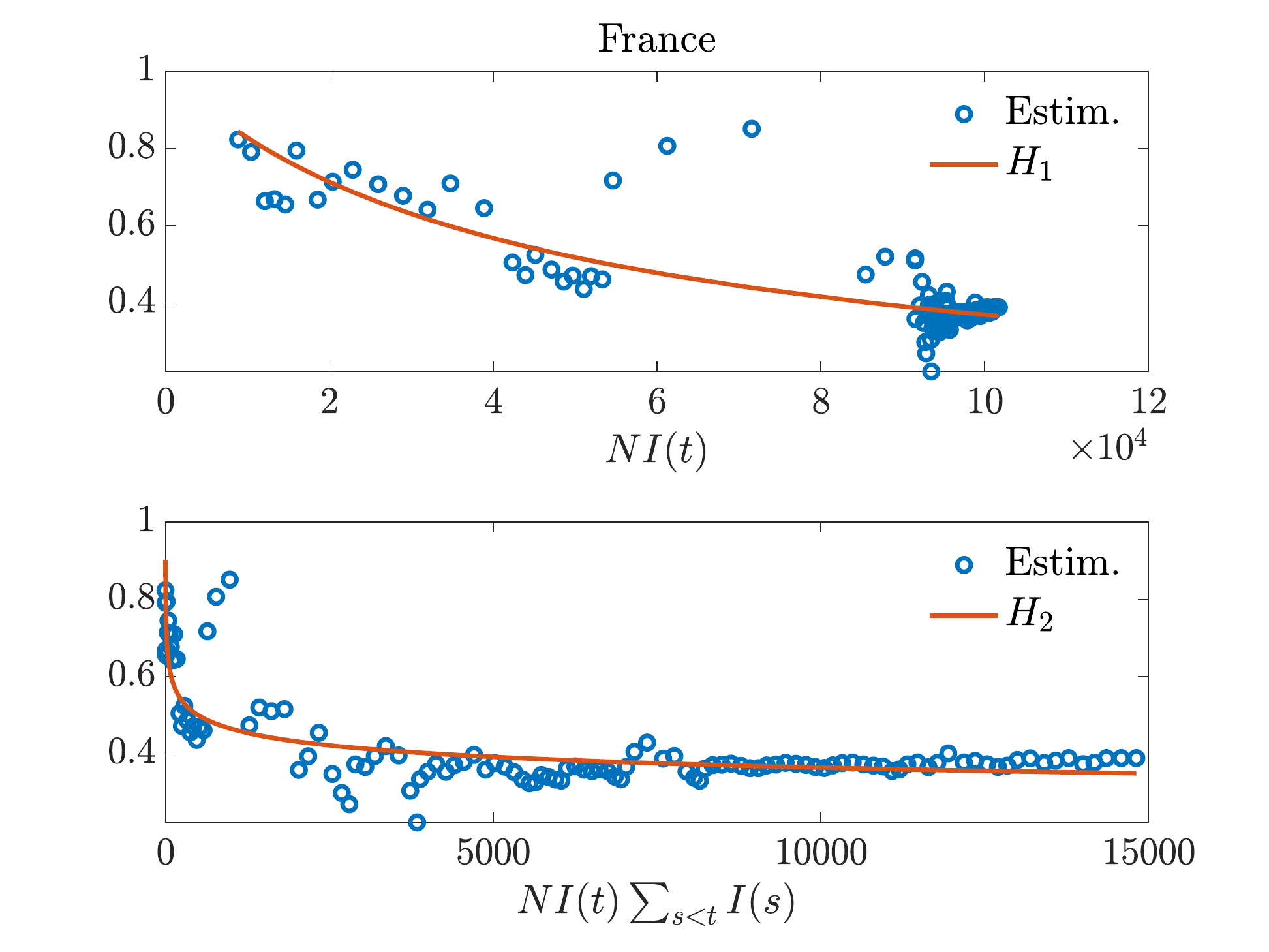}\\
	\includegraphics[scale = 0.4]{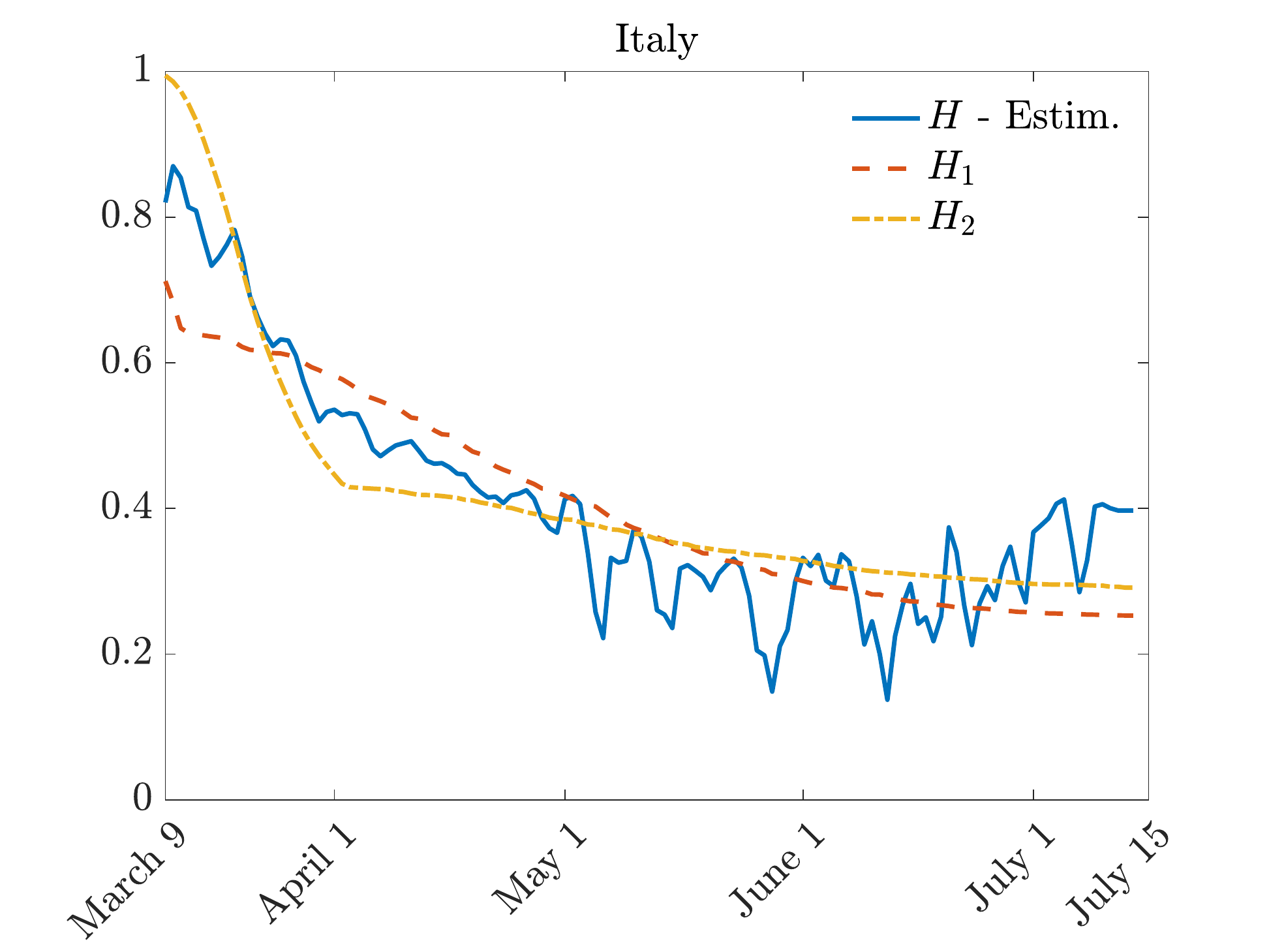}
	\includegraphics[scale = 0.4]{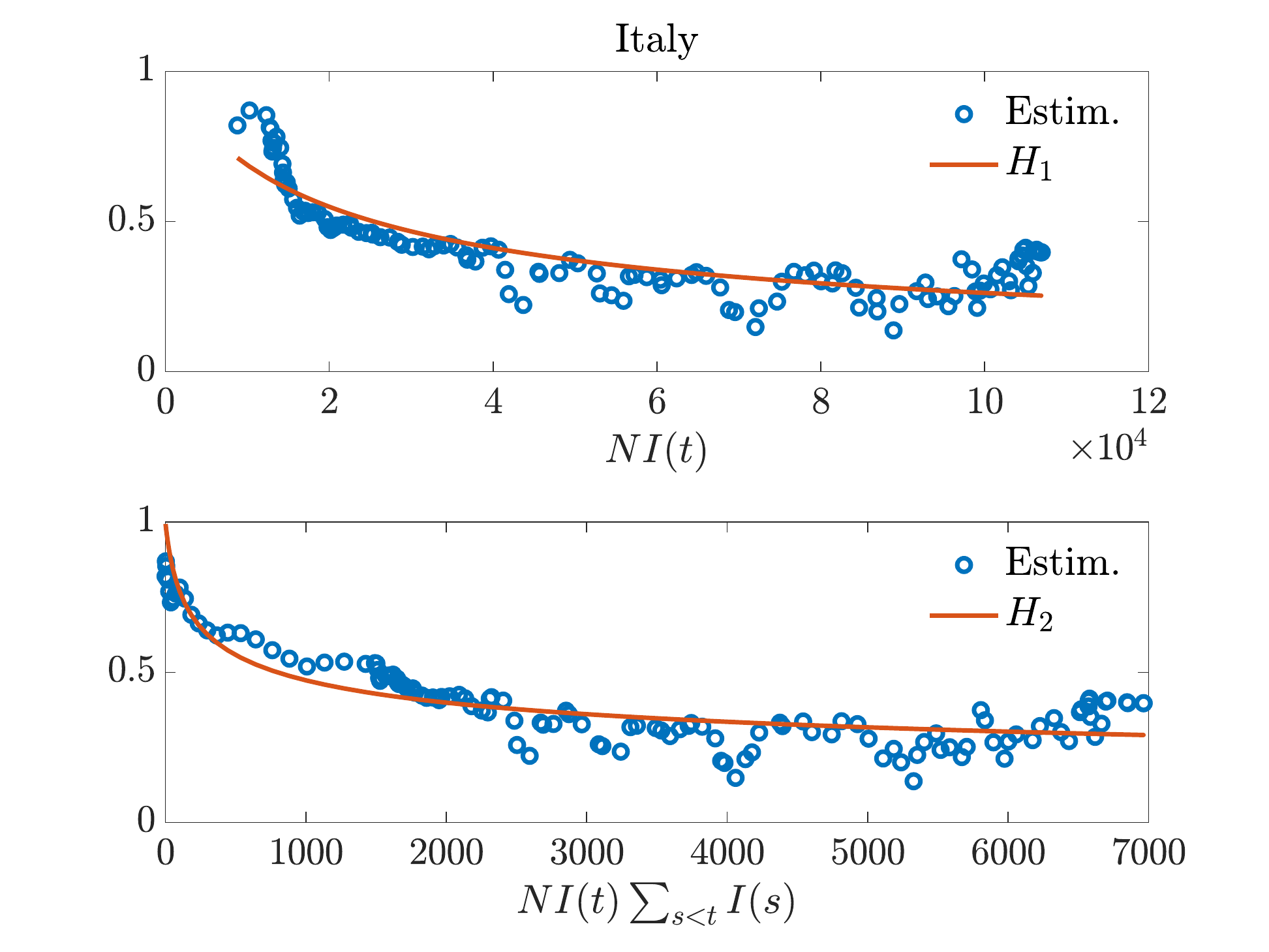}\\
	\includegraphics[scale = 0.4]{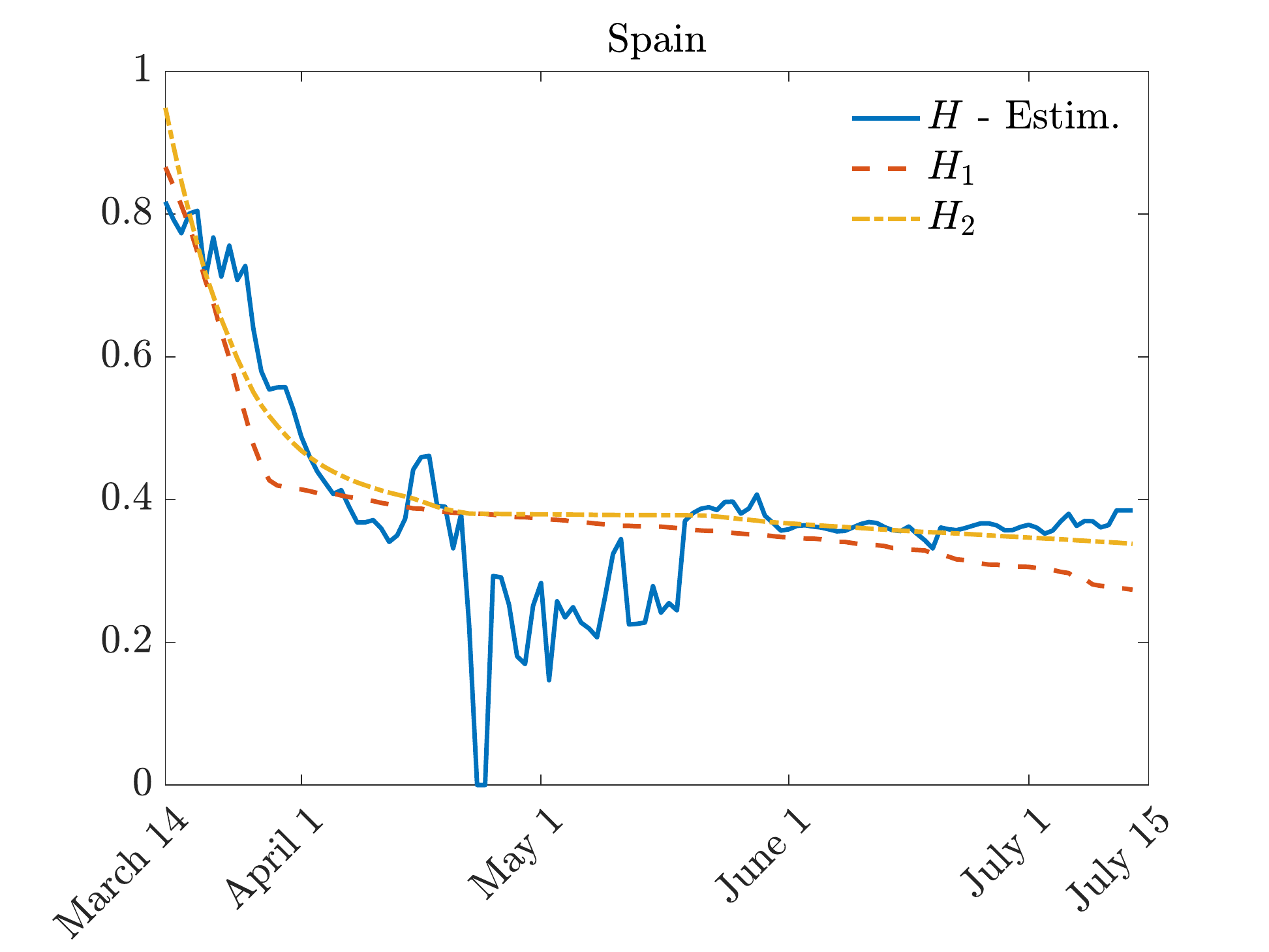}
	\includegraphics[scale = 0.4]{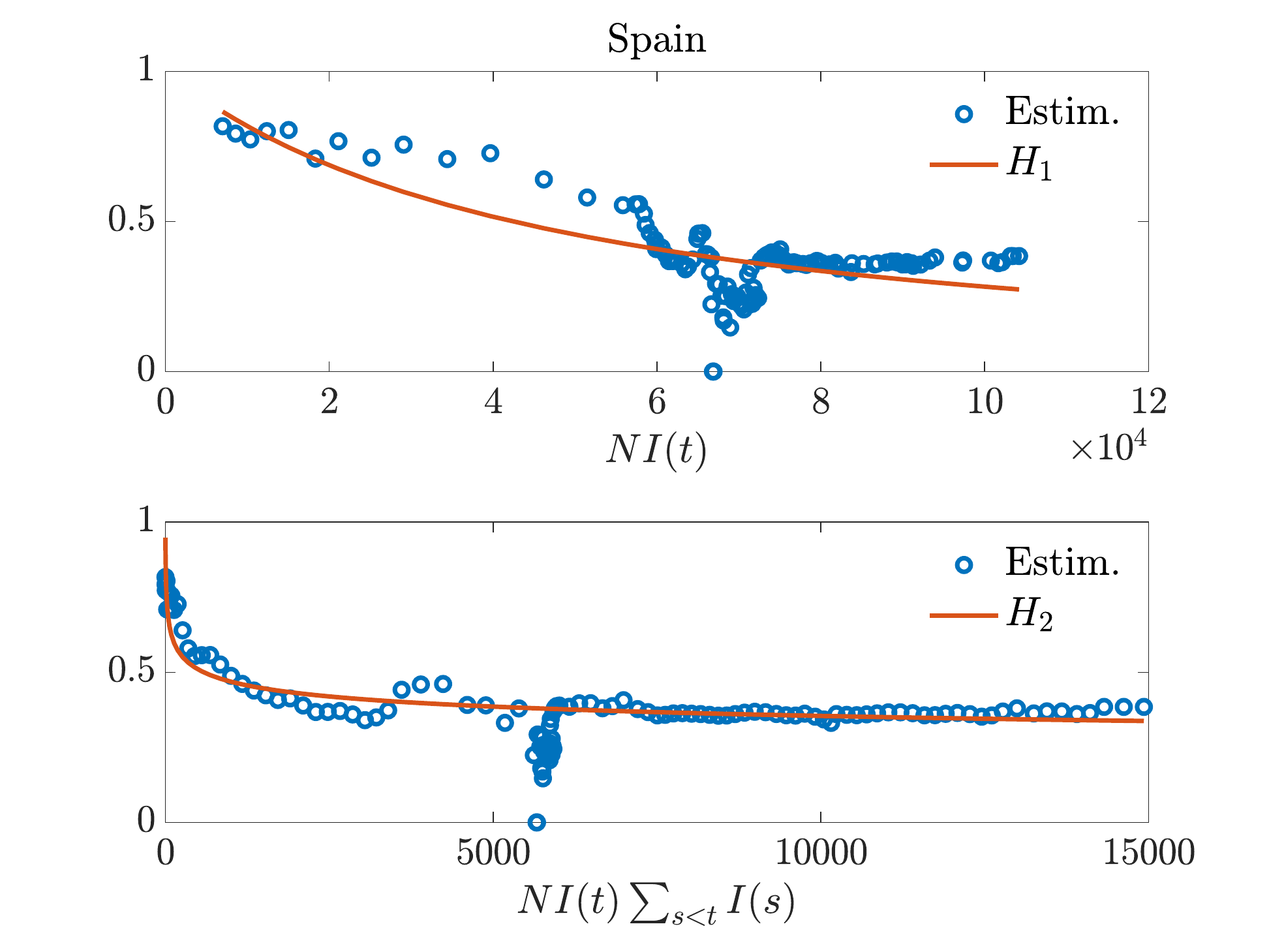}
	\caption{\textbf{Test 3}. Estimated shape of the function $H$ in several European countries (left plots) and its dependency on the variables  $NI(t)$ and $N I (t) \sum_{s<t}I(s)$ (right plots). }
	\label{fig:t3_2}
\end{figure}

In Figure~\ref{fig:t3_1}, we present the result of such fitting procedure between the model \eqref{sir-mass-closed} and the experimental data. The evolution of the estimated $H(t)$, $t \in [t_\ell, T]$, is instead presented in the left column of Figure~\ref{fig:t3_2}. From this figure, it can be observed in the case of Italy how, even if the daily number of infected decreases after May 1st, the estimated $H$ remains quite stable after this day. This behavior cannot be reproduced by using a function $H(I(t))$ as the one given in \eqref{ese}. Instead, a function $H(t,I(t))$ of the form \eqref{ese1}, which takes into account both the instantaneous number of confirmed infections $N I(t)$ and the total number of infected in the population $N \sum_{s<t} I(s)$ up to time $t$, is able to better better fit the observed results. In fact, it appears that taking into account possible memory effects on the contact function is a good path to follow. In the right column of Figure~\ref{fig:t3_1}, the two discussed contact functions have been denoted with $H_1 = H_1(NI(t))$ and $H_2 = H_2(t,NI(t) \sum_{s<t}I(s))$. Both are assumed of the form
\[
H_i(r) = \dfrac{1}{(1+ar)^b}, \qquad i = 1,2.
\]
A least square minimization approach has been finally developed to determine $a>0, b \in \mathbb R$ that best fit the estimated curve whose results are presented in Table \ref{tab:H12}. We can easily observe in the right column of Figure~\ref{fig:t3_2} how the function $H_2$ is capable to better explain the estimated values of $H$ especially after the epidemic peak. To evaluate the goodness of fit we can finally use the so-called coefficient $R^2$, as reported in Table \ref{tab:H12}. Results show that the function $H_2$ appears more suitable in terms of the fitting results for all tested situations. This fact may indicate that people are rather fast to apply social distancing, and therefore to reduce their average number of contacts, whereas they tend to restore the pre-pandemic average contact rate more slowly, possibly due to further memory effects. 

\begin{table}[t]
	\begin{center}
		\begin{tabular}{c | c | c |  c }
			$H_1$ & $\underset{\textrm{Mar 18 - Jul 15}}{\textrm{France}}$& $\underset{\textrm{Mar 10 - Jul 15}}{\textrm{Italy}}$ & $\underset{\textrm{Mar 15 - Jul 15}}{\textrm{Spain}}$  \\ 
			\hline\hline
			& & & \\[-.3cm]
			$a$  & $3 \cdot 10^{-5}$ & $9.038 \cdot 10^{-5}$& $ 1.749 \cdot 10^{-5}$ \\
			\hline
			$b$ & $0.7168$ & $0.5808$ & $1.249$ \\
			\hline
			$R^2$ & $0.7096$ & $0.7542$ & $0.5795$  \\
			\hline \hline
			$H_2$ & $\underset{\textrm{Mar 18 - Jul 15}}{\textrm{France}}$& $\underset{\textrm{Mar 10 - Jul 15}}{\textrm{Italy}}$ & $\underset{\textrm{Mar 15 - Jul 15}}{\textrm{Spain}}$  \\ 
			\hline\hline
			& & & \\[-.3cm]
			$a$  & $1.41$ & $0.01744$ & $  0.5201$ \\
			\hline
			$b$ & $0.1055$ & $0.2566$ & $0.1211$ \\
			\hline
			$R^2$ & $0.7007$ & $0.7959$ & $0.9899$  
		\end{tabular}
	\end{center}
	\caption{Fitting parameters for the estimate of the contact functions $H_i$, $i=1,2$ in different countries based on the evolution of $H(t)$ solution of the optimisation problem \eqref{eq:min_probl2}. The corresponding $R^2$ coefficient is also reported.  }
	\label{tab:H12}
\end{table}

\subsection{Test 4: S-SIR model with fitted contact function}\label{numericsIV}
In this last part, we discuss the results of the S-SIR model when the contact function has the shape extrapolated in the previous paragraph. In particular, we devoted it to show whether the obtained extrapolated function $H$, which depends on the product between the current number of infected and the total number of infected, produces qualitative trends that are in agreement with the data. 

As discussed in Section \ref{numericsIII}, it appears that the curve of infection may be better explained in the case in which the contact function is both a function of the instantaneous number of infected and of the total number of people that contracted the disease up to time $t$. In order to compare qualitatively the observed curve of infected and the theoretical one, we consider the following setting for the three countries under study: $\nu=4, \delta=8$ and $x_J=15$, $\Delta t=0.01$, $\tau=0.01$, $M=10^5$. Moreover, we suppose $S(t=0)$ and $I(t=0)$ to match the relative number of susceptible and infected of each country at the time in which we start our comparison. Finally, we consider
\be
H\left(t,NI(t)\sum_{s<t}I(s)\right)=\frac{1}{(1+aNI(t)\sum_{s<t}I(s))^{b}}, 
\ee
where we use the parameters of Table \ref{tab:H12} that we recall here for the seek of clarity: $(a,b)=(1.41,0.15)$ in the case of France, and $(a,b)=(0.017,0.25)$ in the case of Italy. The case of Spain will be discussed later.

\begin{figure}
	\centering
	\includegraphics[scale = 0.5]{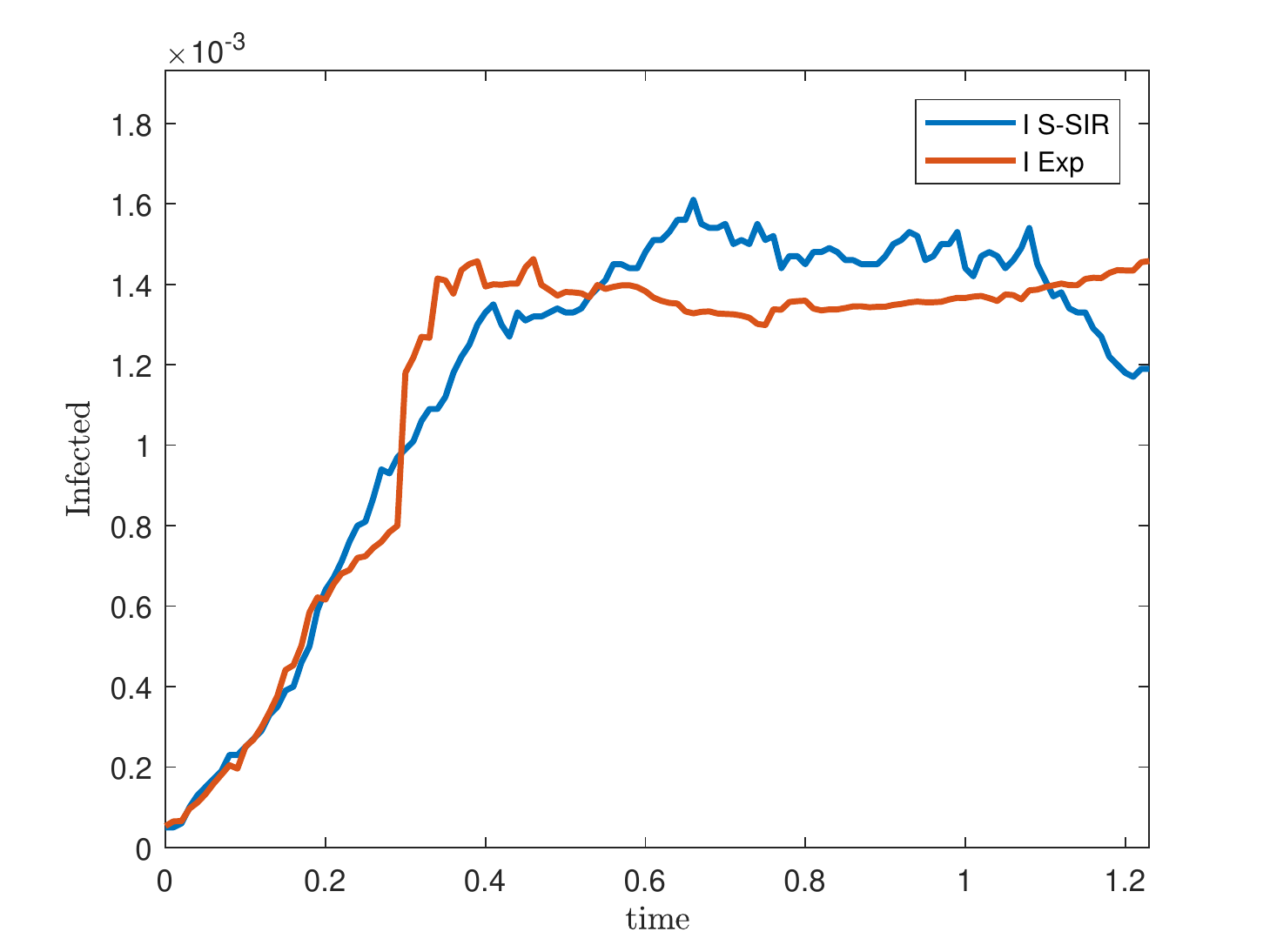}
		\includegraphics[scale = 0.5]{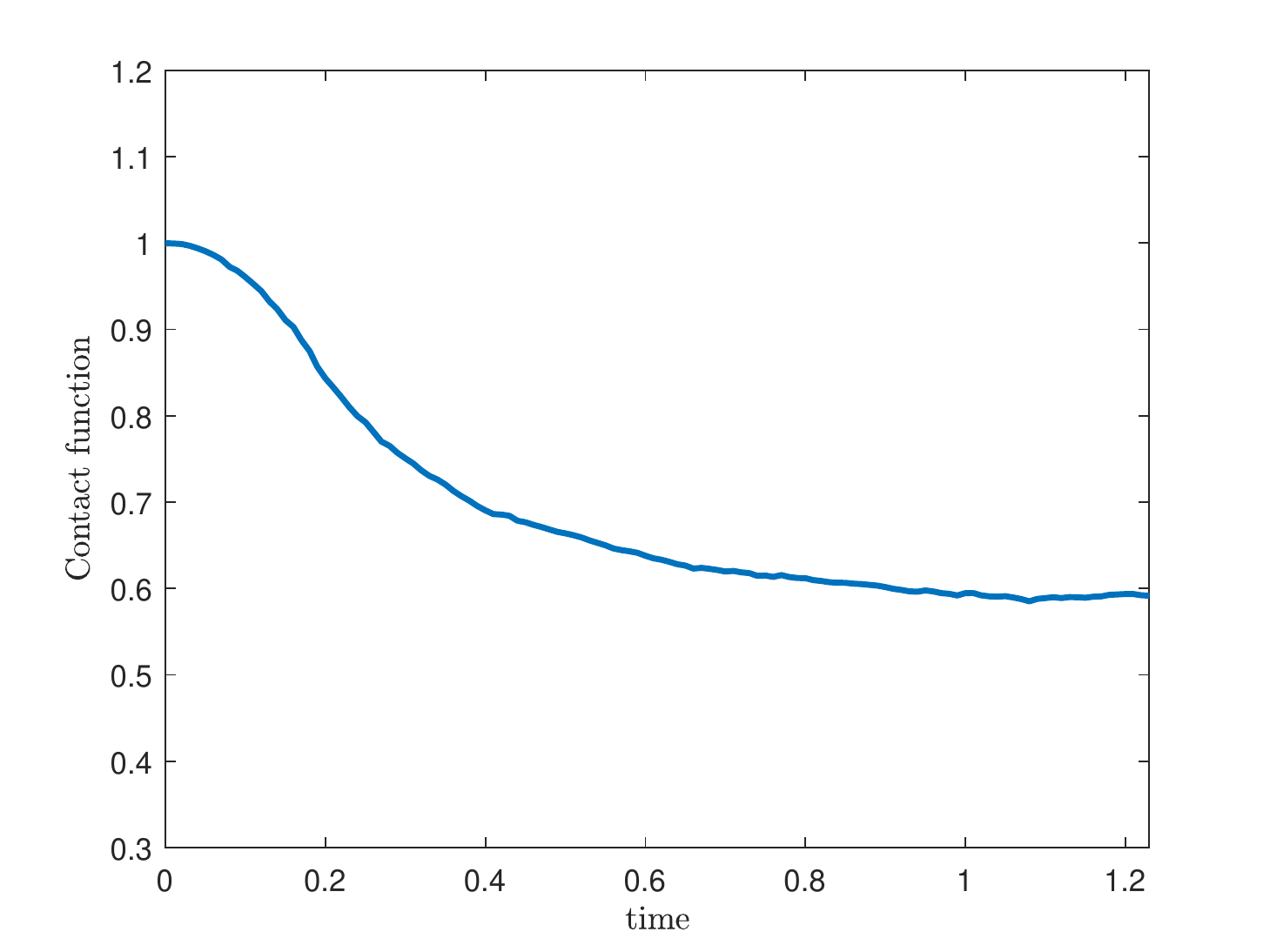}\\
	\includegraphics[scale = 0.5]{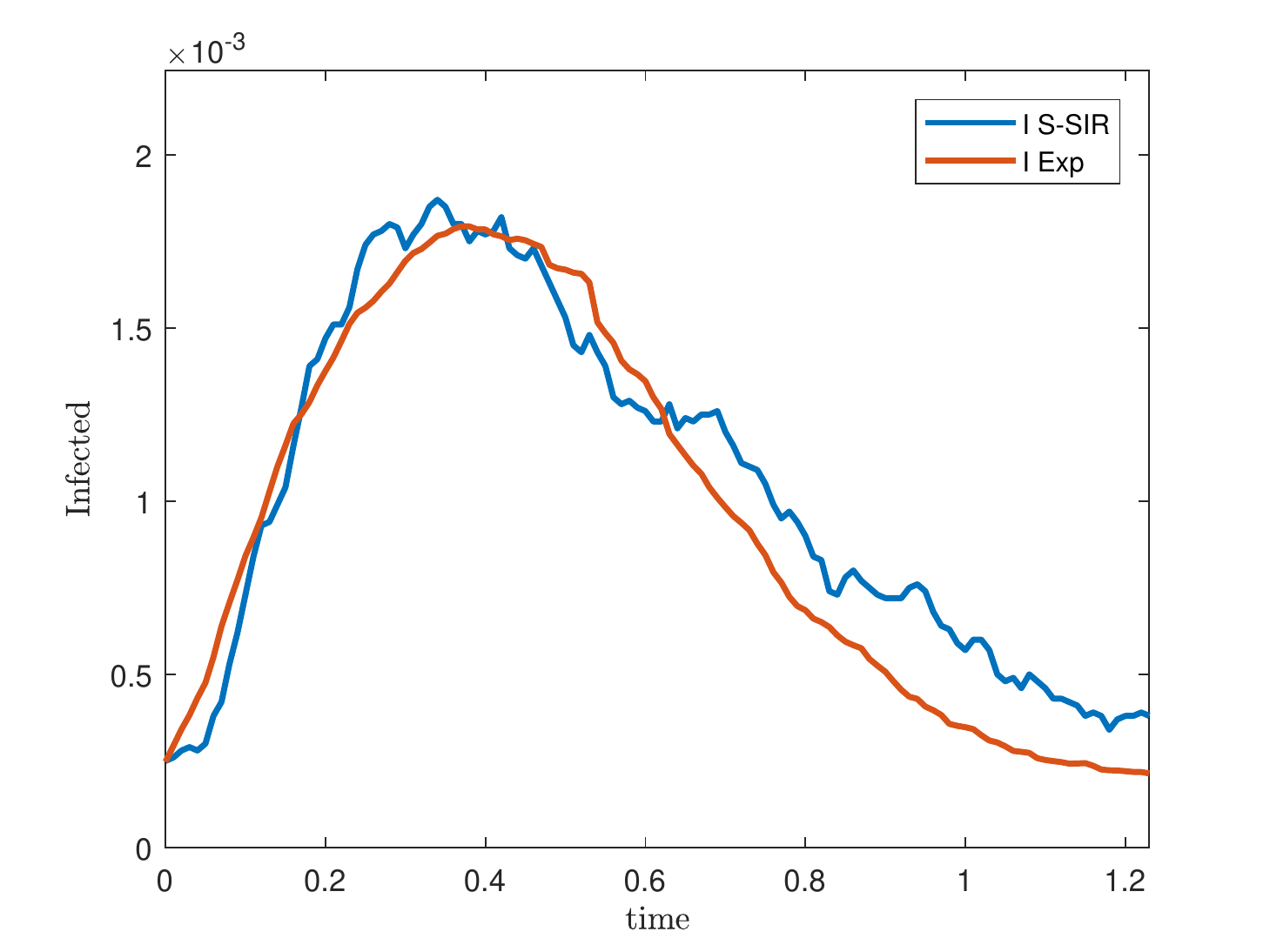}
	\includegraphics[scale = 0.5]{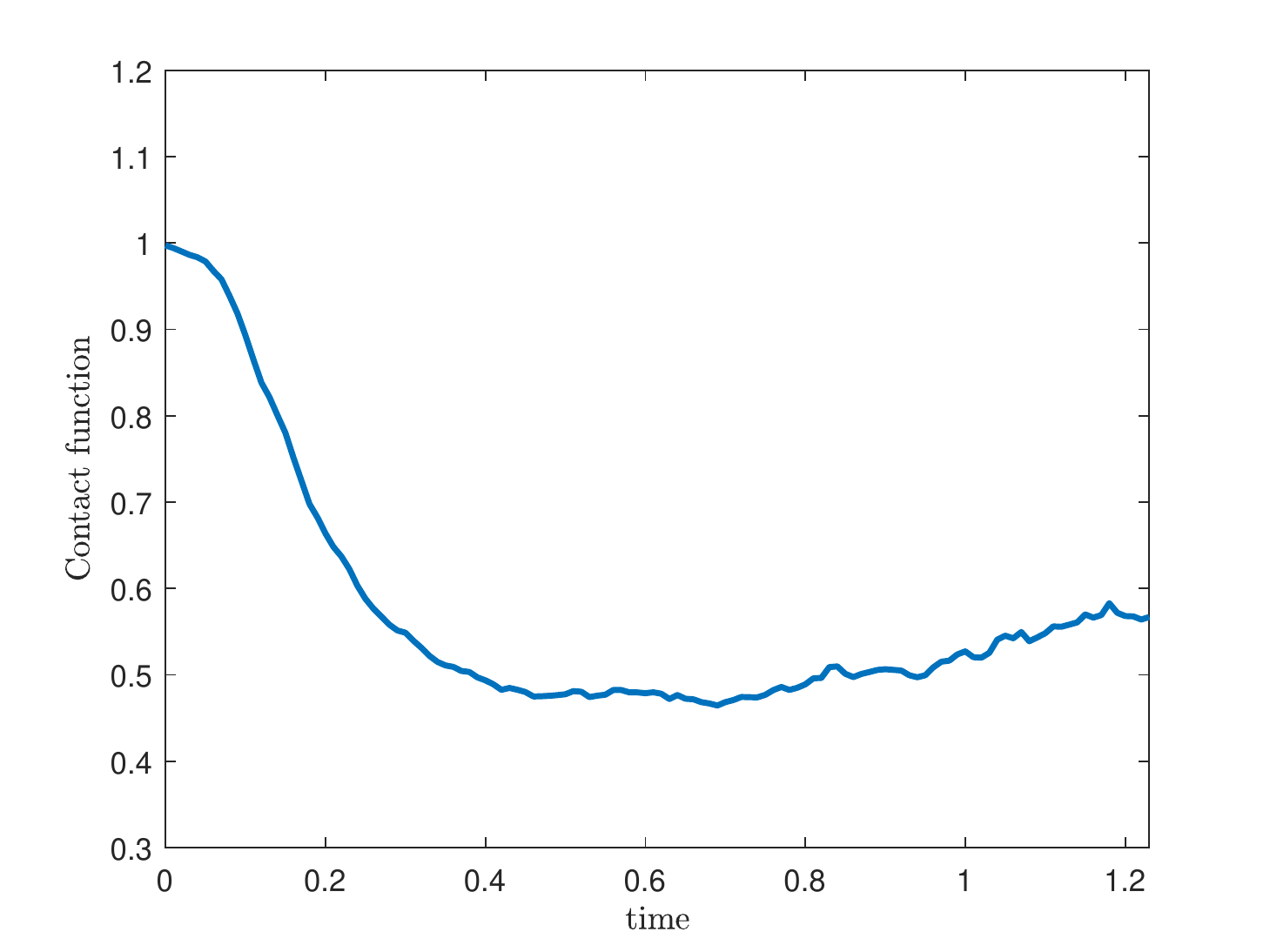}\\
	\caption{\textbf{Test 4}. Left: relative number of infected over time for the S-SIR model when memory effects are taken into account in the contact function. Right: corresponding contact function over time. Top: the French case. Bottom: the Italy case. }
	\label{fig:t4_1}
\end{figure}

In Figure~\ref{fig:t4_1} we show the profiles of the infected over time together with the shape of the function $H$ again over time. The results show that with the choices done for the contact function, it is possible to reproduce at least qualitatively the shape of the trend of infected during the pandemic observed in Italy and in France.

\begin{figure}
	\centering
	\includegraphics[scale = 0.5]{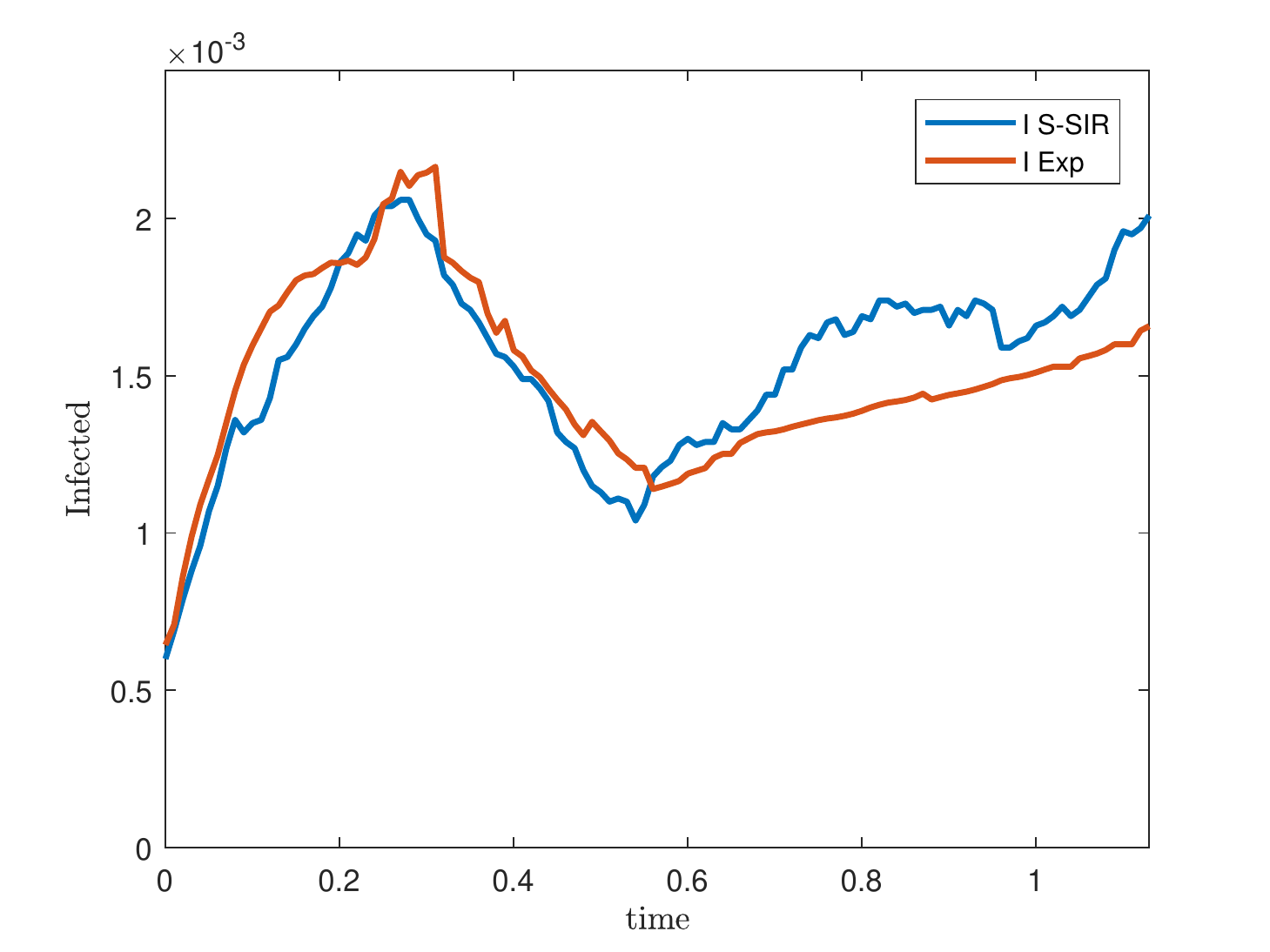}
	\includegraphics[scale = 0.5]{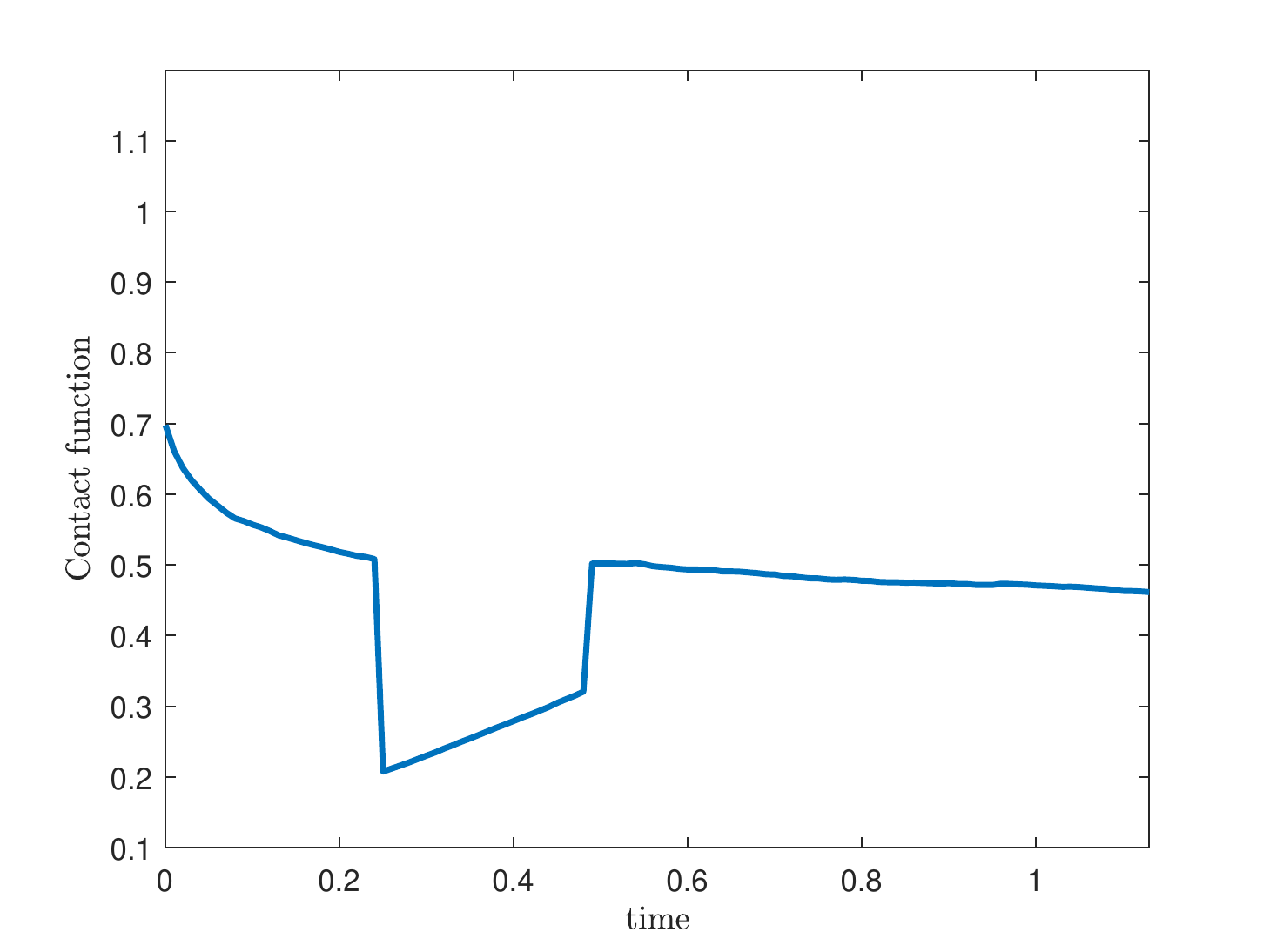}\\
	\caption{\textbf{Test 4}. Left: relative number of infected over time for the S-SIR model when memory effects are taken into account in the contact function. Right: corresponding contact function over time. Spain case.}
	\label{fig:t4_2}
\end{figure}

It is worth to remark that the considered social parameters have been estimated only the in the case of France, see \cite{Plos}, and we assumed that the initial contact distribution is the same for the Italian case. 
We now consider the case of Spain. For this country, according to Figure \ref{fig:t3_1}, the trend of infected undergoes a deceleration during the lockdown period. This can be also clearly observed in Figure \ref{fig:t3_2} where the extrapolated shape of the contact function $H$ is shown. Let also observe that while the global behavior of this function is captured by the fitting procedure, we however lose the minimum which takes place around end of April. This minimum is responsible of the deceleration in the number of infected and can be brought back to a strong external intervention in the lifestyle of Spain country with the scope of reducing the hospitalizations. This effect can be reproduced by our model by imposing the same behavior in the function $H$. To that aim, the Figure \ref{fig:t4_2} reports finally the profile of the infected over time together with the shape of the function $H$ again over time for this last case. The results show that also in this case, the S-SIR model is capable to qualitatively reproduce the data.  

\section*{Conclusions}
The development of strategies for mitigating the spreading of a pandemic is an important  public health priority. The recent case of COVID-19 pandemic has seen as main strategy  restrictive measures on the social contacts of  the population, obtained by household quarantine, school or workplace closure, restrictions on travels, and, ultimately, a total lockdown.  Mathematical models represent powerful tools for a better understanding of this complex landscape of intervention strategies and for a precise quantification of the relationships \rev{between} potential costs and benefits of different options \cite{Ferg}. In this direction, we introduced a system of kinetic equations coupling the distribution of social contacts with the spreading of a pandemic driven by the rules of the SIR model, aiming to explicitly quantify the mitigation  of the pandemic in terms of the reduction of the number of social contacts of individuals.
The kinetic modeling of the statistical distribution of social contacts has been developed according to the recent results in \cite{Plos}, which present an exhaustive description of contacts in the France population, divided by categories. The numerical experiments then show that the kinetic system is able to capture most of the phenomena related to the effects of partial lockdown strategies, and, eventually to maintain pandemic under control.

\section*{Acknowledgement} This work has been written within the
activities of GNFM group  of INdAM (National Institute of
High Mathematics), and partially supported by  MIUR project ``Optimal mass
transportation, geometrical and functional inequalities with applications''.
The research was partially supported by
the Italian Ministry of Education, University and Research (MIUR): Dipartimenti
di Eccellenza Program (2018--2022) - Dept. of Mathematics ``F.
Casorati'', University of Pavia. 
G.D. would like to thank the Italian Ministry of Instruction, University and Research (MIUR) to support this research with funds coming from PRIN Project 2017 (No. 2017KKJP4X entitled “Innovative numerical methods for evolutionary partial differential equations and applications”).
B.P. has received funding from the European Research Council (ERC) under the European Union's Horizon 2020 research and innovation program (grant agreement No 740623).


\end{document}